\documentclass[cameraready]{Interspeech}

\newcommand{\issl}{\textsc{insideSSL}}
\title{\issl: Understanding Self-Supervised\\ Speech Representations using a Model-Centric Perspective}

\author[affiliation={1}]{Samir}{Sadok}
\author[affiliation={1}]{Xavier}{Alameda-Pineda}

\address{
    $^1$ Inria, Université Grenoble Alpes CNRS, LJK, France
}

\email{samir.sadok@inria.fr, xavier.alameda-pineda@inria.fr}

\keywords{Self-supervised learning, speech representation, entropy, curvature, invariance, interpretability.}

\usepackage{comment}
\usepackage{subcaption}
\usepackage{multirow}
\graphicspath{{../images/}}
\usepackage{amsfonts}       
\usepackage[dvipsnames]{xcolor}
\usepackage[most]{tcolorbox}
\usepackage{color, colortbl, arydshln}
\usepackage{adjustbox}
\usepackage{pifont}

\definecolor{modernblue}{HTML}{0056B3}

\definecolor{burgundy}{HTML}{922B21}
\definecolor{steelblue}{HTML}{008080}
\definecolor{midnightblue}{HTML}{191970} 
\newcommand{\ssl}[1]{\textcolor{midnightblue}{\small\scshape #1}}

\begin{document}

\maketitle


\begin{abstract}
Self-supervised learning (SSL) models, such as Wav2Vec2, HuBERT, and WavLM, have become foundational across a wide range of speech and audio tasks. Despite their success, understanding their internal layer-wise dynamics remains an ongoing challenge. To address this, we propose a two-part model-centric framework called \issl. First, we establish a task-agnostic analysis from three intrinsic per-layer perspectives: compression (entropy), geometry (curvature), and robustness to perturbations. We show that varying training objectives induce distinct regimes of acoustic compression and manifold unfolding. Second, we introduce the cross-layer Generative Compatibility Matrix (GCM) to evaluate functional transferability, exposing stable phonetic cores, identity volatility, and deep-layer semantic pruning. In addition to these evaluations, linear probing connects the model-centric perspective to downstream tasks, demonstrating how layer topology dictates phoneme, pitch, and speaker encoding.
\end{abstract}



\section{Introduction}
Self-supervised speech representation learning (SSL) has become a cornerstone of modern audio processing \cite{liu2022audio, mohamed2022self}, with models such as \ssl{WavLM} \cite{chen2022wavlm}, \ssl{Wav2Vec2} \cite{baevski2020wav2vec}, and \ssl{HuBERT} \cite{hsu2021HuBERT} achieving remarkable performance across speech recognition \cite{baevski2020wav2vec, chen2022wavlm}, speaker verification \cite{fan2020exploring, chen2022wavlm, wang2021fine}, emotion recognition \cite{pepino2021emotion, chen2023exploring, wang2021fine} and speech enhancement tasks \cite{hung2022boosting, huang2022investigating}. By leveraging unlabeled audio data, these models learn representations that capture meaningful semantic and acoustic information without relying on explicit supervision. However, despite their empirical success, the internal representations learned by SSL speech models are still not fully understood. Understanding how these models encode and compress information is crucial. This knowledge also informs their geometric organization, interpretability, and downstream performance.

\begin{figure}[t!]
    \centering
    \includegraphics[width=1\linewidth]{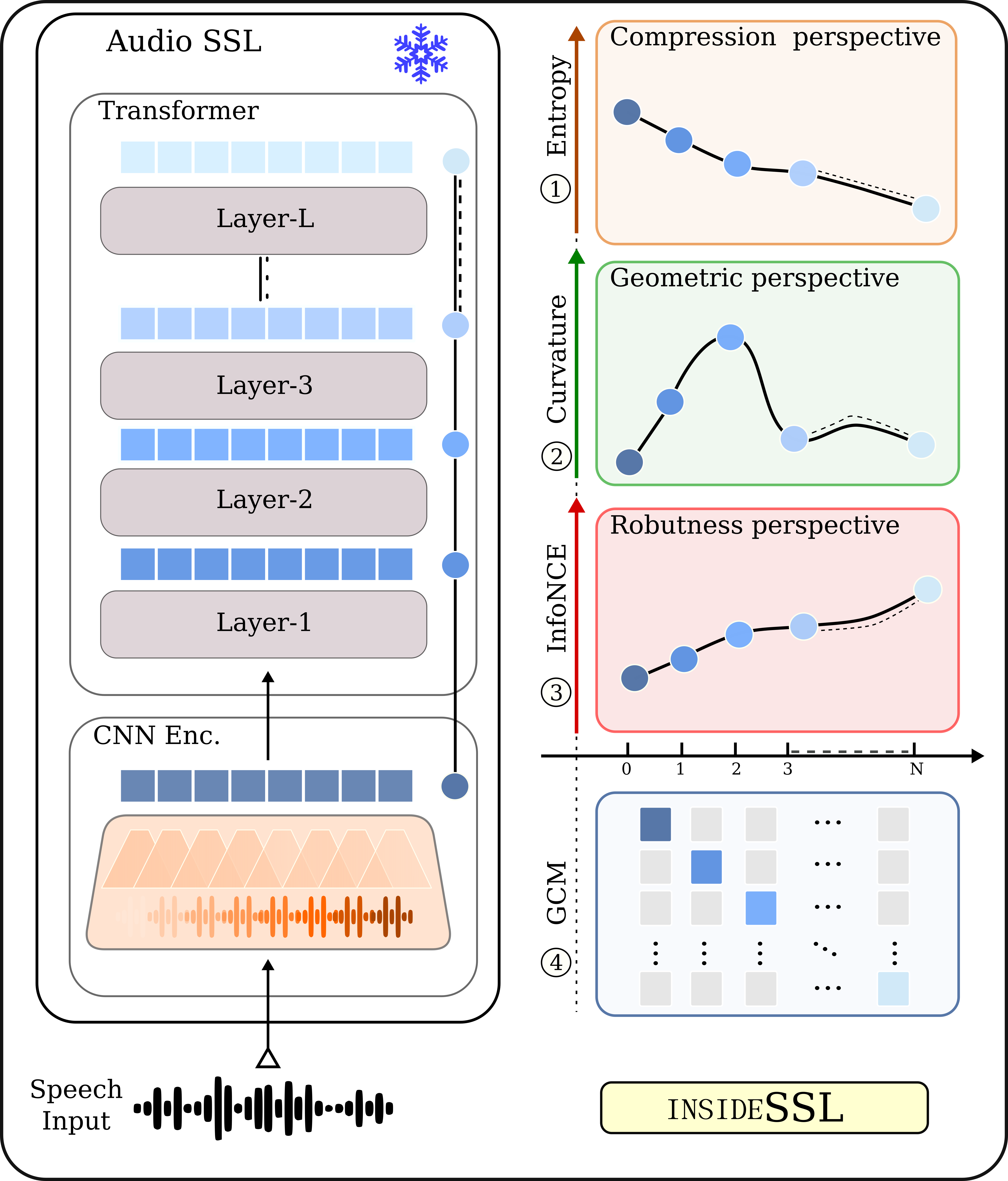}
    \caption{Systematic layer-wise evaluation of SSL speech representations using a model-centric perspective framework. It tracks the evolution of entropy, curvature, and invariance across layer depth to characterize how information is compressed, organized, and abstracted within the Transformer hierarchy.}
    \label{fig:inside-SSL}
\end{figure}

Unlike most prior studies \cite{fan2020exploring, pasad2021layer, shah2021all, riera2023phone, li2023exploration, ashihara2024self, pasad2023comparative, chiu2025probing, chiu2025large} that analyze SSL models primarily through their correlation with predefined speech attributes usually linked to downstream tasks, we propose \issl\footnote{{\color{burgundy}Project page and code:} \href{https://samsad35.github.io/audio-ssl-dynamics-site/}{\texttt{https://insideSSL.github.io/}}} (see Figure~\ref{fig:inside-SSL}), a model-centric task-agnostic framework for the analysis of speech SSL models, consisting in per-layer and cross-layer tools. Instead of relying solely on external labels and inspired by recent analyses of representation dynamics in language models \cite{skean2025layer}, we first investigate the intrinsic, task-agnostic properties of the per-layer embeddings along three complementary perspectives, namely compression, geometry, and robustness:

\noindent\emph{(i) Compression perspective:} Guided by the Information Bottleneck principle \cite{tishby2000information, shwartz2017opening, shwartz2024compress, skean2025layer}, we investigate the trade-off between representational diversity and signal compression. Using matrix-based von Neumann entropy, we quantify the \emph{informational density} of the embeddings. This allows us to track how the effective dimensionality evolves across layers, clearly distinguishing between early states of high-dimensional spread and deep regimes of information compression.

\noindent\emph{(ii) Geometry perspective:} Inspired by recent work on high-dimensional embeddings \cite{hosseini2023large, skean2025layer}, we explore the geometric structure of token trajectories. By measuring the average curvature of transitions between adjacent tokens, we examine how representations evolve from capturing abrupt, local acoustic details in early layers to forming smoother, more linear, and abstract structures deeper in the network.

\noindent\emph{(iii) Robustness perspective:} Robustness to input perturbations is a hallmark of meaningful representations. We analyze whether embeddings remain stable under transformations such as additive noise, pitch shifts, or time masking \cite{oord2018representation, thilak2023lidar, skean2023dime}. By using an InfoNCE-based metric to approximate the mutual information between augmented views, we evaluate the exact layer depth at which models maintain invariant representations versus where they become brittle to signal distortions.

To deepen this layer-by-layer understanding, we also introduce the cross-layer Generative Compatibility Matrix (GCM). Each entry of the GCM quantifies how well generative decoders trained on the embeddings of one layer (row index) can interpret those of another layer (column index). The GCM reveals the transferability and structural stability of phonetic content and speaker identity across the network hierarchy. 

In addition to this model-centric and task-agnostic metrics, we explicitly bridge these per-layer and cross-layer perspectives with task-specific probing (e.g., phoneme classification, pitch regression). By linking representation topology to downstream performance, our study sheds light on the internal mechanisms of widely-used models like \ssl{Wav2Vec2}, \ssl{HuBERT}, and \ssl{WavLM}, offering valuable insights that can inform the design of more interpretable and task-aligned speech architectures, hence the name \issl.


\section{\issl's Methodology}
\label{sec:methodology}

\subsection{Overview}
In this work, we analyze self-supervised speech models by examining their token embeddings across all layers. Our model-centric analysis focuses on three complementary per-layer perspectives (compression, geometry, and robustness) and the cross-layer generative compatibility matrix.
For a given input signal $\mathbf{x}$, we denote the corresponding token embeddings at layer $\ell$ by a matrix $\mathbf{Z}^{(\ell)} \in \mathbb{R}^{N \times D}$, where $N$ is the number of tokens and $D$ the embedding dimension. We define the layer index $\ell \in \{0, \dots, L\}$, where $\mathbf{Z}^{(0)}$ represents the output of the CNN feature extractor (serving as input to the Transformer backbone) and $\mathbf{Z}^{(L)}$ corresponds to the final Transformer layer output, as illustrated in Figure~\ref{fig:inside-SSL}.
Each row $\mathbf{z}_n^{(\ell)}$ represents the embedding of the $n$-th token at layer $\ell$. These embeddings serve as the foundation for all subsequent analyses.
The metrics presented in the following are computed per-sample $\mathbf{x}$, and then averaged over the test or evaluation sets.

\subsection{Per-layer: Compression, Geometry, and Robustness}
\label{sec:invariance}
The per-layer analysis is performed via three metrics: \emph{compression} (how embeddings compress or preserve semantic information), \emph{geometry} (how embeddings are organized in high-dimensional space), and \emph{robustness} (how invariant embeddings are to input perturbations), that are described in the following.

\noindent\textbf{Compression (Entropy).}
To quantify the amount of information preserved across layers, we use the matrix-based entropy~\cite{yu2019multivariate, skean2025layer}. Given the Gram matrix $\mathbf{K}^{(\ell)} = \mathbf{Z}^{(\ell)} (\mathbf{Z}^{(\ell)})^\top$, we define the normalized eigenvalues as $\tilde{\lambda}_r = \frac{\lambda_r\left(\mathbf{K}^{(\ell)}\right)}{\mathrm{tr}(\mathbf{K}^{(\ell)})}$, where $\lambda_r\left(\mathbf{K}^{(\ell)}\right)$ are the eigenvalues. The entropy is then defined as the von Neumann entropy:
\begin{equation}
E^{(\ell)}(\mathbf{x}) = -\sum_{r=1}^{R} \tilde{\lambda}_r \log \tilde{\lambda}_r,
\end{equation}
where $R = \mathrm{rank}(\mathbf{K}^{(\ell)})$. This measure effectively quantifies the spectral spread of the data representation without requiring explicit probability density estimation.
High entropy values indicate that embeddings are spread across many dimensions (high diversity), whereas low entropy values suggest that the representation lies in a low-dimensional subspace, signaling strong information compression.\\

\noindent\textbf{Geometric Structure (Curvature).}
Following~\cite{hosseini2023large, skean2025layer}, we characterize the embedding manifold by the curvature of token trajectories. Let $\mathbf{v}_n^{(l)} = \mathbf{z}_{n+1}^{(\ell)} - \mathbf{z}_n^{(\ell)}$ be the transition vector, the average curvature is given by: 
\begin{equation}C^{(\ell)}(\mathbf{x}) = \frac{1}{N-1} \sum_{n=1}^{N-1} \arccos \left(\frac{(\mathbf{v}_{n+1}^{(\ell)})^\top \mathbf{v}_n^{(\ell)}}{\parallel\mathbf{v}_{n+1}^{(\ell)}\parallel ~ \parallel\mathbf{v}_n^{(\ell)}\parallel}\right).
\end{equation}
High curvature reflects abrupt changes typical of local acoustic or phonetic details. Conversely, a decrease in curvature indicates smoother, more linear trajectories, often associated with the abstraction of global, semantically coherent structures. By tracking curvature across layers, we gain insight into how SSL models transition from encoding local acoustic details to representing higher-level semantic information.\\

\noindent\textbf{Robustness (InfoNCE).}
To assess robustness, we generate two augmented views for each sample  denoted as $\mathbf{x}^A$ and $\mathbf{x}^B$, by applying perturbations such as noise, masking, or pitch shifting. We evaluate the alignment of their representations at layer $\ell$ using the InfoNCE loss~\cite{oord2018representation}. For a given token $n$, the embeddings $(\mathbf{z}^{(\ell)}_{n,A}, \mathbf{z}^{(\ell)}_{n,B})$ constitute the positive pair, while the embeddings of all other samples in the batch serve as negatives:

\begin{equation}
  I_{\tau}^{(\ell)} = - \frac{1}{N} \sum_{n=1}^N \log \frac{\exp\left(\text{sim}(\mathbf{z}^{(\ell)}_{n,A}, \mathbf{z}^{(\ell)}_{n,B}) / \tau\right)}{\sum_{k \neq n} \exp\left(\text{sim}(\mathbf{z}^{(\ell)}_{n,A}, \mathbf{z}^{(\ell)}_{k, B}) / \tau\right)},
\end{equation}

where $\text{sim}(\cdot, \cdot)$ denotes cosine similarity and $\tau$ is the temperature parameter, set to $0.1$ in our experiments. Minimizing this objective is equivalent to maximizing a lower bound on the mutual information between the latent representations of the invariant views. Consequently, a lower $I_{\tau}^{(\ell)}$ implies high invariance of the $\ell$-th layer to the introduced perturbations, whereas increasing scores in deeper layers reveal that the representations become brittle to signal distortions.

\subsection{Cross-layer: Generative Compatibility Matrix (GCM)}
\label{subsec:clpm}
In addition to the per-layer tools described in the previous section, InsideSSL also analyses the links between those layers.

\noindent\textbf{General principle.~} We introduce the \textit{Generative Compatibility Matrix} (GCM) to systematically map the representational topology of self-supervised speech models. Grounded in a \emph{generative reconstruction paradigm}, this framework learns mappings from SSL latent representations at any given depth back to the input space or to low-level features (e.g., Mel-spectrograms). Crucially, this represents the first methodology to explicitly quantify \emph{cross-layer functional relationships} in audio SSL. By shifting focus from the isolated layer-wise evaluations to inter-layer dependencies, our approach elucidates the precise trajectory of information flow, revealing how phonetic content and speaker identity are dynamically encoded, abstracted, or preserved throughout the network hierarchy.

Specifically, for each layer $\ell \in \{0, \dots, L\}$ of the SSL model, we train a decoder $D^{(\ell)}$ (denoted as the \textit{model layer}) to reconstruct the target representation ($\mathbf{x}$) by minimizing the generative loss conditioned on the representations $\mathbf{Z}^{(\ell)}$. During the evaluation phase, we perform cross-layer inference: the decoder $D^{(\ell)}$, optimized for layer $\ell$, is conditioned on representations $\mathbf{Z}^{(k)}$ from a different layer $k$ (denoted as the \textit{evaluation layer}). The resulting matrix is defined as:

$$ \text{GCM}^\mathcal{M}(\ell, k) = \mathcal{M}\left(D^{(\ell)}(\mathbf{Z}^{(k)}), \mathbf{x}\right),$$
where $\mathcal{M}$ represents a performance metric such as the L1 Loss, and effectively measure the compatibility between layers from a generative standpoint.

Illustrative Example: Consider a decoder $D^{(3)}$ trained on the representations of layer $3$. If this decoder maintains strong performance when evaluated on representations from layer $6$ (i.e., a favorable $\text{GCM}^\mathcal{M}(3,6)$ value, whether high or low depending on the metric), it implies that the representation of layer $6$ remains sufficiently similar to that of layer $3$ to be interpreted by the same decoding function. Conversely, a degradation in performance would indicate a representational shift in the functional alignment between these depths.

The methodology associated with \issl~covers both the per-layer (compression, geometry, robustness) and the cross-layer (generative compatibility) metrics. We evaluate these in the next section, and complete the analysis with standard linear probing experiments. 

\section{Experiments}
\subsection{Per-Layer Analysis: Experimental Setup}
\label{subsec:setup}
We evaluate our model-centric perspective on several widely used SSL models. For each model, we extract hidden representations at every layer and analyze them according to the three perspectives described in Section~\ref{sec:methodology}.\

\noindent\textbf{Models.} All self-supervised learning models examined in this study utilize a bidirectional Transformer backbone, yet their pre-training objectives differ fundamentally (see Table~\ref{tab:model_summary} for an overview).
\ssl{Wav2Vec2} \cite{baevski2020wav2vec} learns contextualized representations via a contrastive task, requiring the model to identify the correct quantized latent speech unit for a masked time step among a set of distractors.
\ssl{HuBERT} \cite{hsu2021HuBERT} adopts a similar masked prediction paradigm but relies on an offline clustering step to generate discrete pseudo-labels, enabling iterative refinement of targets.
\ssl{WavLM} \cite{chen2022wavlm} extends the \ssl{HuBERT} framework by integrating a \textit{masked denoising modeling} objective. Here, inputs are corrupted with noise or overlapping speech, and the model is trained to predict the pseudo-labels of the original clean signal, thereby enhancing robustness in complex acoustic environments.
Finally, \ssl{Data2Vec-audio} \cite{baevski2022data2vec} diverges with a modality-agnostic teacher--student framework. Instead of predicting discrete units, it regresses the continuous latent representations averaged from the teacher's top layers, offering a purely continuous pathway to self-supervision.
Beyond architectural objectives, these models are categorized by scale and data volume. The {\scshape base} configurations serve as the standard reference, while {\scshape plus} variants (e.g., \ssl{WavLM-base-plus}) typically retain the base architecture but leverage significantly larger training datasets. In contrast, {\scshape large} models scale up the capacity by expanding both network depth and embedding dimensions. Unless otherwise specified, we refer to the {\scshape base} configuration by default.

\begin{table}[t!]
\caption{Overview of investigated SSL architectures by scale and objective. \textsc{base} serve as the reference; \textsc{plus} denotes expanded training data to 94k hours, and \textsc{large} scales depth to 24 layers. Models are categorized by their primary learning task: ``P'' for predictive (masked prediction), ``C'' for contrastive, and ``D'' for denoising.\label{tab:model_summary}}%
\renewcommand{\arraystretch}{1.4}%
\setlength{\tabcolsep}{4pt}%
\resizebox{\columnwidth}{!}{%
\begin{tabular}{ccccccc}
\toprule
\multirow{2.5}{*}{\rotatebox{0}{\textbf{Scale}}} 
& \multirow{2.5}{*}{\textbf{\shortstack{SSL\\Model}}} 
& \multirow{2.5}{*}{\textbf{\shortstack{Pretraining\\Dataset}}} 
& \multicolumn{3}{c}{\textbf{Task}} 
& \multirow{2.5}{*}{\textbf{\shortstack{Architecture\\Details}}} \\ 
\cmidrule{4-6}
& & 
& \rotatebox{0}{P} 
& \rotatebox{0}{C} 
& \rotatebox{0}{D} 
& \\
\midrule
\multirow{5}{*}{\rotatebox{0}{{\scshape base}}} 
& \ssl{Wav2Vec2}  & \multirow{5}{*}{\shortstack{Librispeech\\(960h)}} & \ding{55} & \checkmark & \ding{55} & \multirow{7}{*}{\shortstack{\# Layers: 12\\Hidden Dim: 768\\Model size: 95M}} \\ 
& \ssl{HuBERT} & & \checkmark & \ding{55} &  \ding{55} & \\
& \ssl{WavLM} &  & \checkmark &  \ding{55} & \checkmark & \\
& \ssl{Data2vec} &  & \checkmark & \ding{55}  & \ding{55} & \\
& \ssl{UniSpeech} &  & \checkmark &  \checkmark & \ding{55} & \\
\cmidrule{1-6}
\rotatebox{0}{{\scshape plus}}
& \ssl{WavLM} & \multirow{1}{*}{\shortstack{Mix (94kh)}}  & \checkmark & \ding{55} & \checkmark & \\
\midrule
\multirow{5}{*}{\rotatebox{0}{{\scshape large}}} 
& \ssl{Wav2Vec2} & \multirow{5}{*}{\shortstack{Mix\\($\geq$60kh)}} & \ding{55} & \checkmark & \ding{55} & \multirow{5}{*}{\shortstack{\# Layers: 24\\Hidden dim: 1024\\Model size: 315M}} \\
& \ssl{HuBERT} & & \checkmark & \ding{55} &  \ding{55} & \\
& \ssl{WavLM} &  & \checkmark &  \ding{55} & \checkmark & \\
& \ssl{Data2vec} &  & \checkmark & \ding{55}  & \ding{55} & \\
& \ssl{UniSpeech} &  & \checkmark &  \checkmark & \ding{55} & \\
\bottomrule
\multicolumn{7}{l}{ \textit{Note:} h: hours; kh: thousand hours; M: million parameters.}
\end{tabular}%
}%
\end{table}

\begin{figure*}[t]
    \centering
    \begin{subfigure}[b]{0.32\textwidth}
        \centering
        \includegraphics[width=\textwidth]{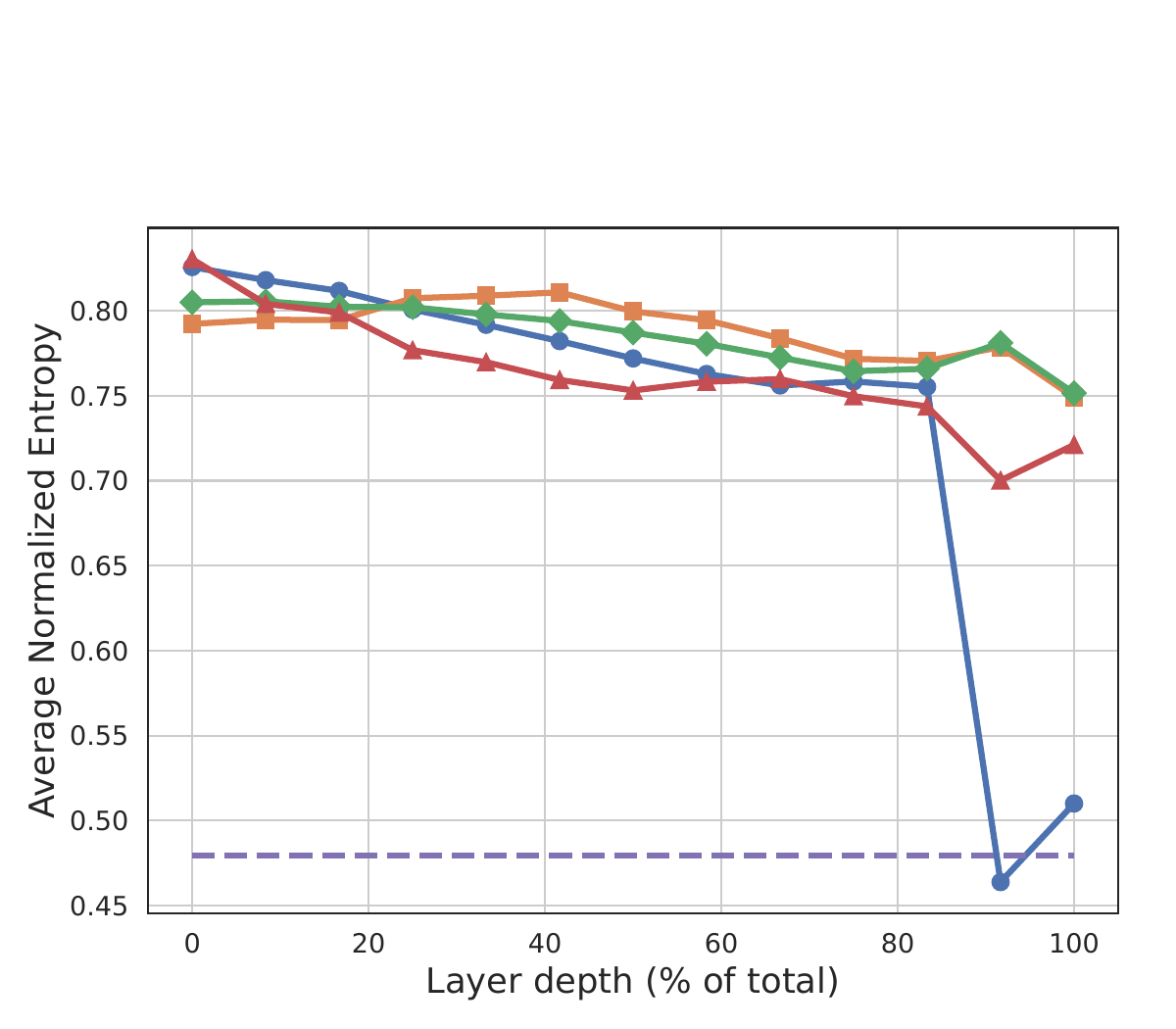}
        \caption{Compression (entropy)}
        \label{fig:fig1a}
    \end{subfigure}
    \hfill
    \begin{subfigure}[b]{0.32\textwidth}
        \centering
        \includegraphics[width=\textwidth]{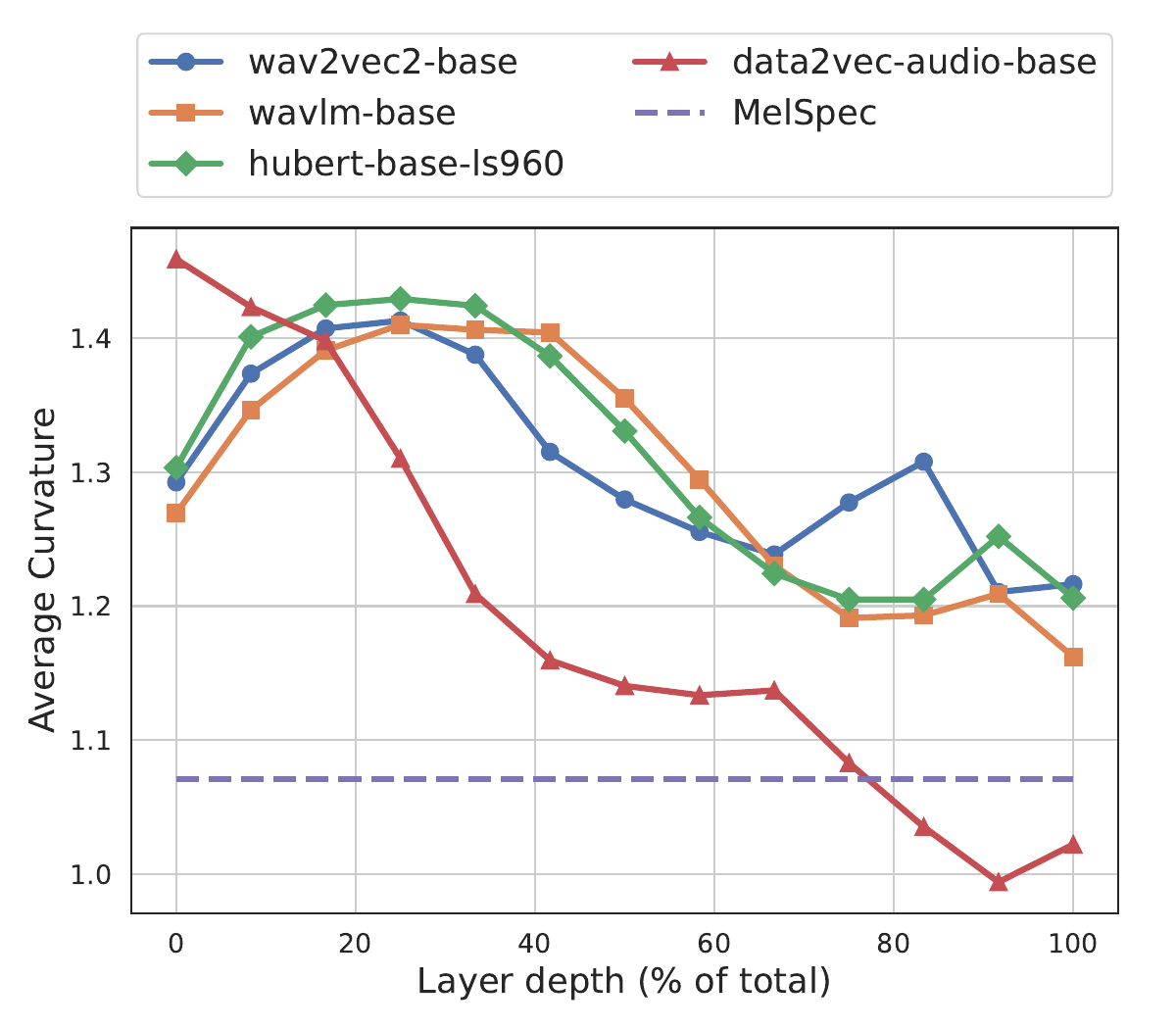}
        \caption{Geometry (curvature)}
        \label{fig:fig1b}
    \end{subfigure}
    \hfill
    \begin{subfigure}[b]{0.32\textwidth}
        \centering
        \includegraphics[width=\textwidth]{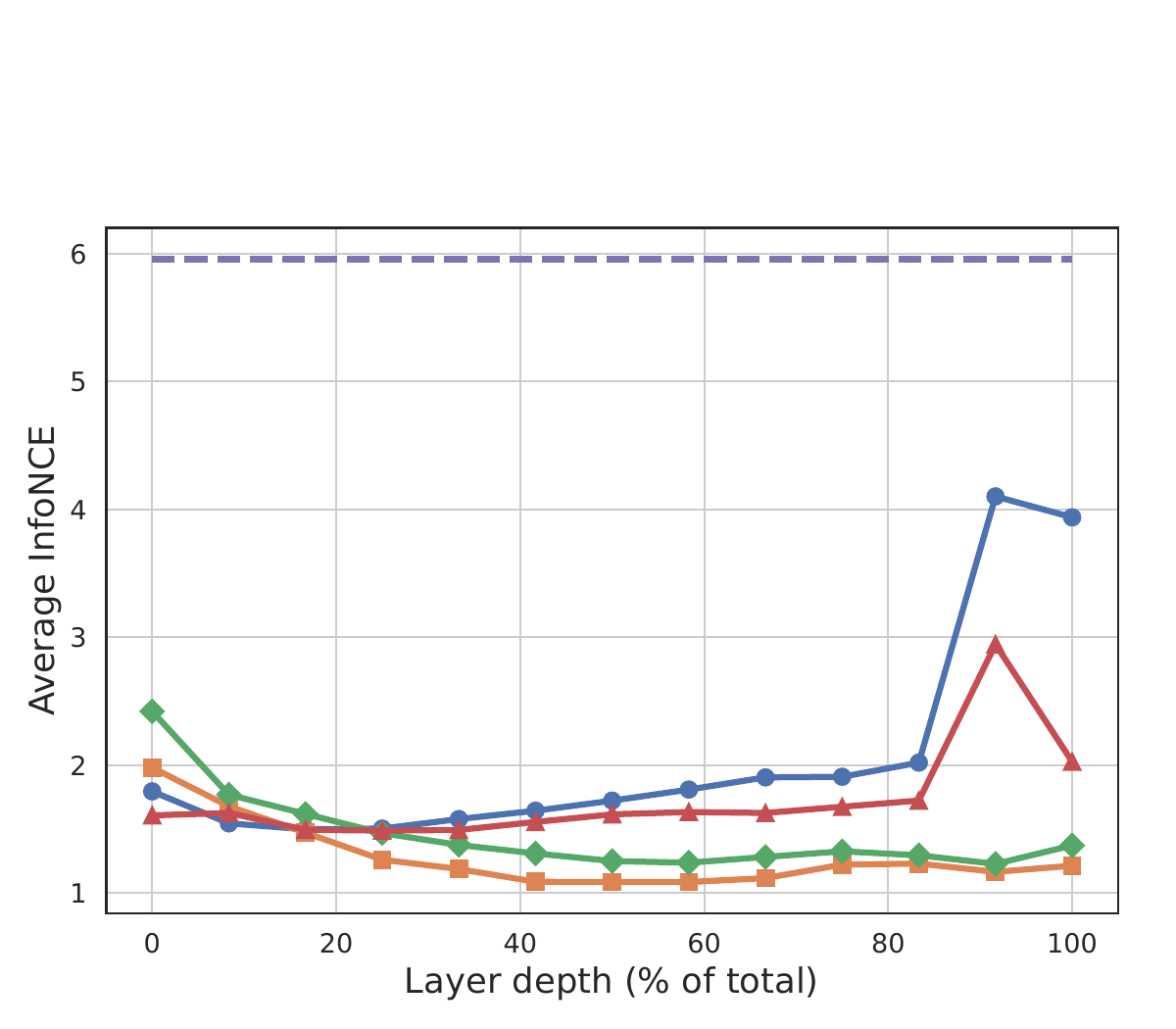}
        \caption{Robustness (InfoNCE)}
        \label{fig:fig1c}
    \end{subfigure}

    \vspace{0.5cm}
    
    \begin{subfigure}[b]{0.32\textwidth}
        \centering
        \includegraphics[width=\textwidth]{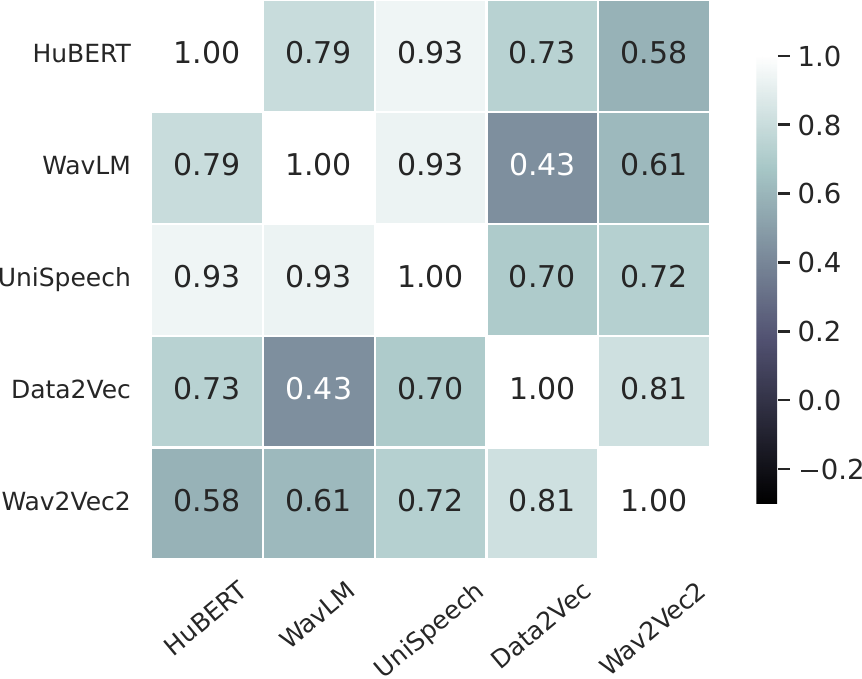}
        \caption{Inter-Model entropy correlation matrix}
        \label{fig:fig1d}
    \end{subfigure}
    \hfill
    \begin{subfigure}[b]{0.32\textwidth}
        \centering
        \includegraphics[width=\textwidth]{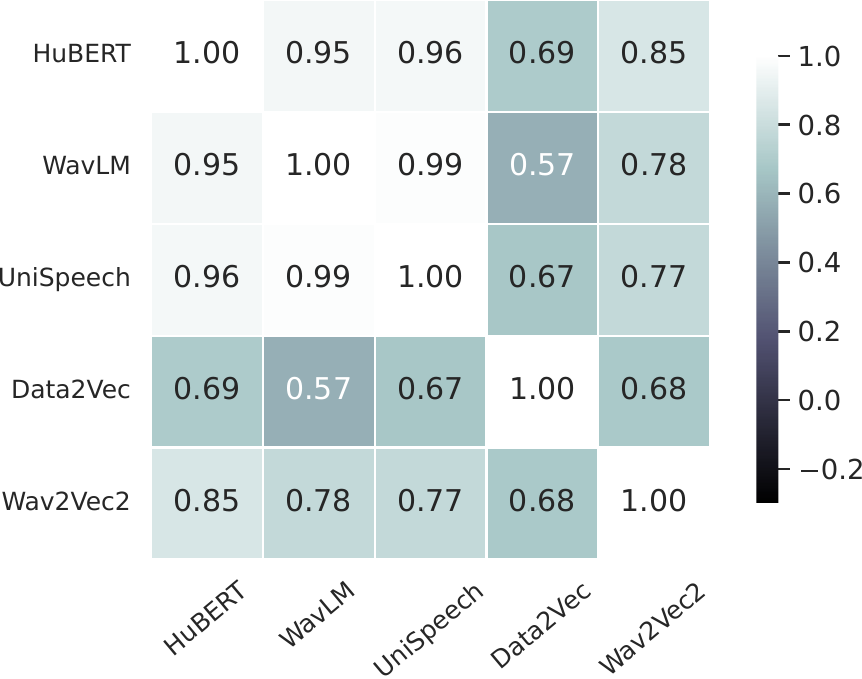}
        \caption{Inter-Model curvature correlation matrix}
        \label{fig:fig1e}
    \end{subfigure}
    \hfill
    \begin{subfigure}[b]{0.32\textwidth}
        \centering
        \includegraphics[width=\textwidth]{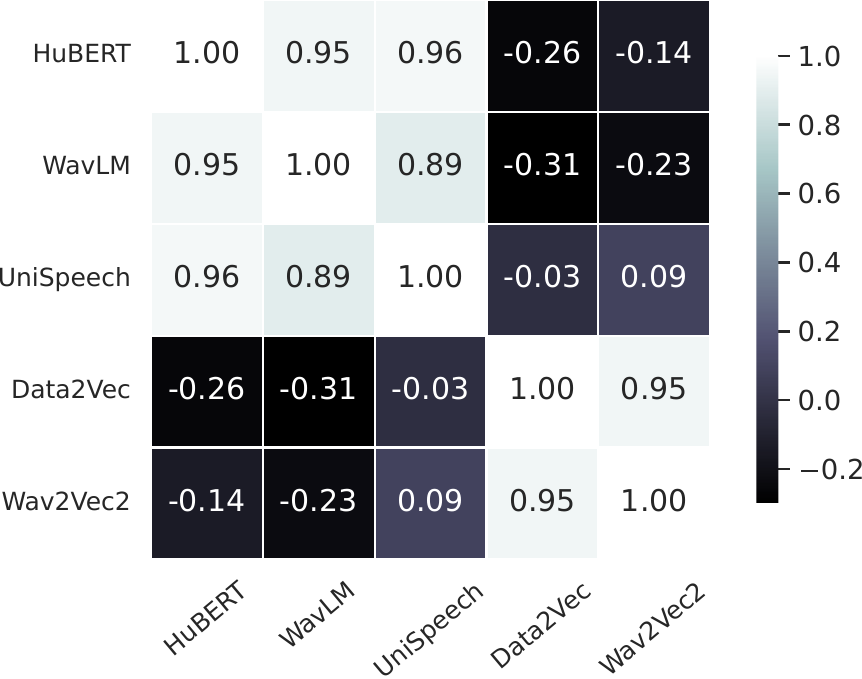}
        \caption{Inter-Model invariance correlation matrix}
        \label{fig:fig1f}
    \end{subfigure}
    
    \caption{Layer-wise analysis of SSL models across the three model-centric perspectives: compression (entropy), geometry (curvature), and robustness (invariance to input perturbations).}
    \label{fig:analysis-3-perspectives}
\end{figure*}

\noindent\textbf{Data.} Unless stated otherwise, all experiments are conducted on the \emph{LibriSpeech} corpus \cite{panayotov2015librispeech}. We use the \emph{test-clean} subset ($2{,}620$ utterances) to compute intrinsic layer-wise metrics (entropy, curvature, and InfoNCE), ensuring consistent and fair comparisons across pre-trained models. The trainable components of our framework—namely the generative decoders used in the cross-layer Generative Compatibility Matrix (GCM; Section~\ref{subsec:clpm}) and the linear task probes (Section~\ref{subsec:task-probing})—are trained on the \emph{train-clean-100} subset.

\noindent\textbf{Augmentations.} To assess invariance (Sec.~\ref{sec:invariance}), we generate two augmented versions of each audio segment using a sequential composition of perturbations. Each transformation within the pipeline is triggered independently with a probability of $p=0.7$. The augmentation chain includes: additive Gaussian noise (amplitude $\in [0.001, 0.015]$), pitch shifting ($\pm 4$ semitones), gain adjustment, and time masking. All augmentations were implemented using the \texttt{audiomentations} library\footnote{\url{https://github.com/iver56/audiomentations}}.
The resulting InfoNCE loss is further normalized by $\log N$ to obtain a bounded mutual-information lower bound, enabling comparison across different batch sizes. 

\noindent\textbf{Implementation details.} All embeddings are computed with the official checkpoints of the corresponding models. The implementation of these intrinsic metrics (entropy, curvature, and InfoNCE) is adapted from the \texttt{information\_flow} \cite{skean2025layer} repository\footnote{\url{https://github.com/OFSkean/information_flow}}. To ensure comparability across different architectures, we normalize the raw entropy by its theoretical maximum. We report results using the \textit{maxEntropy} normalization, where the raw entropy is divided by $\min(\log N, \log D)$, representing the upper bound for a representation of rank $R \leq \min(N, D)$.

\begin{figure*}[t]
    \centering
    \begin{subfigure}[b]{0.32\textwidth}
        \centering
        \includegraphics[width=\textwidth]{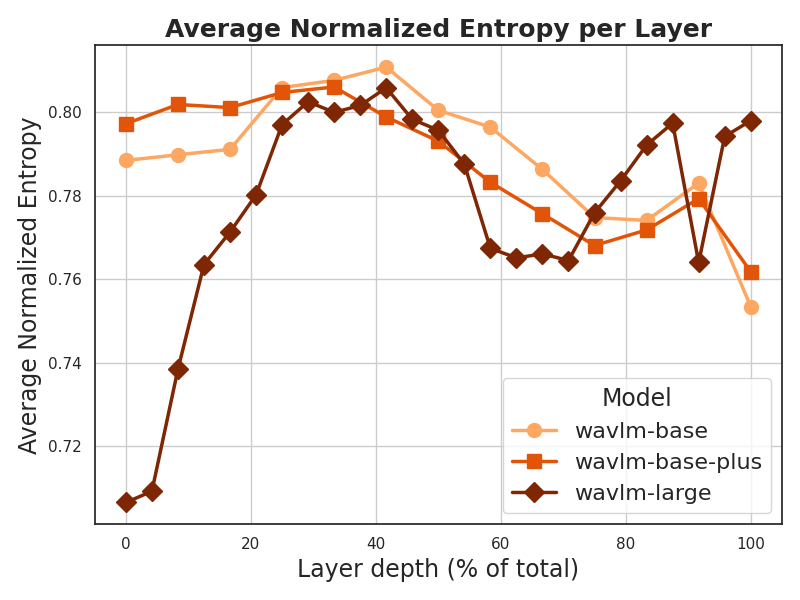}
        \caption{Compression (entropy)}
        \label{fig:fig2a}
    \end{subfigure}
    \hfill
    \begin{subfigure}[b]{0.32\textwidth}
        \centering
        \includegraphics[width=\textwidth]{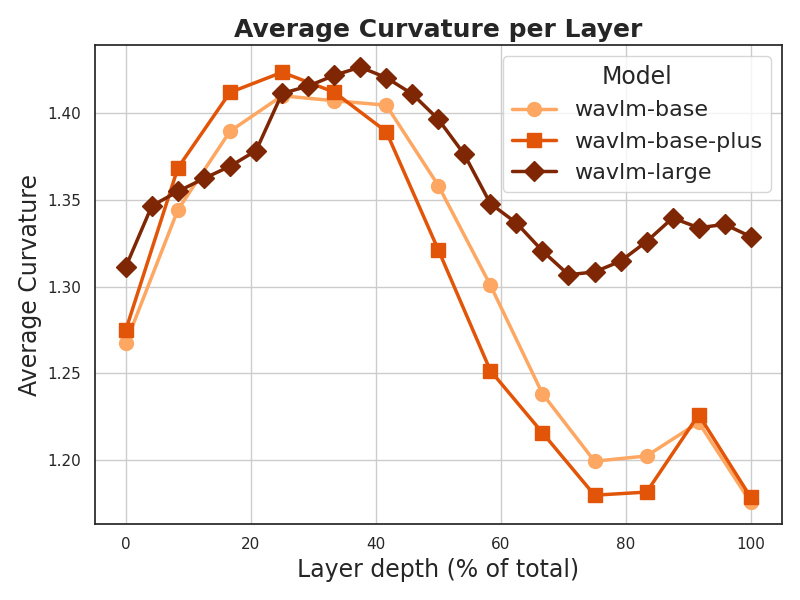}
        \caption{Geometry (curvature)}
        \label{fig:fig2b}
    \end{subfigure}
    \hfill
    \begin{subfigure}[b]{0.32\textwidth}
        \centering
        \includegraphics[width=\textwidth]{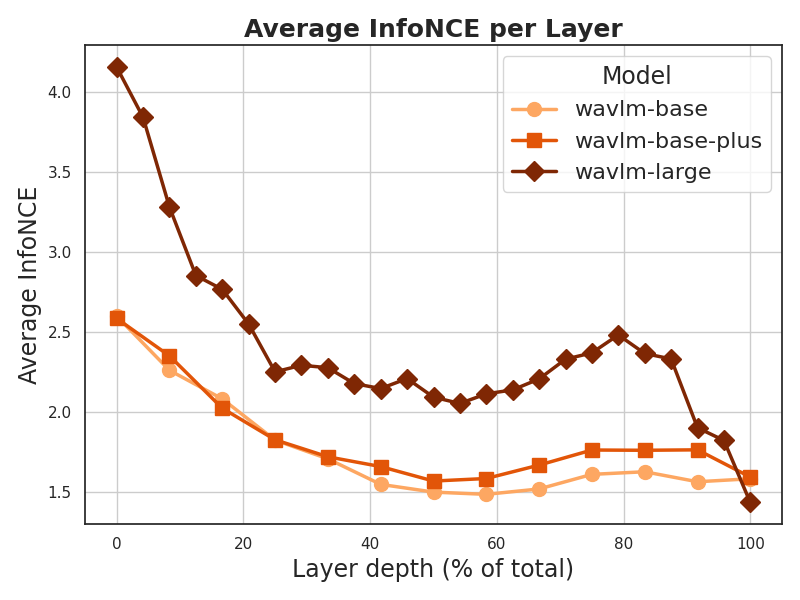}
        \caption{Robustness (InfoNCE)}
        \label{fig:fig2c}
    \end{subfigure}
    
    \caption{The impact of training data (\ssl{WavLM-plus}) and model size (\ssl{WavLM-large}) on SSL audio representations, showing how layer-wise properties (compression, geometry, and robustness) vary across models of different scales.}
    \label{fig:size}
\end{figure*}

\begin{figure*}[t]
    \centering
    \begin{subfigure}[b]{0.32\textwidth}
        \centering
        \includegraphics[width=\textwidth]{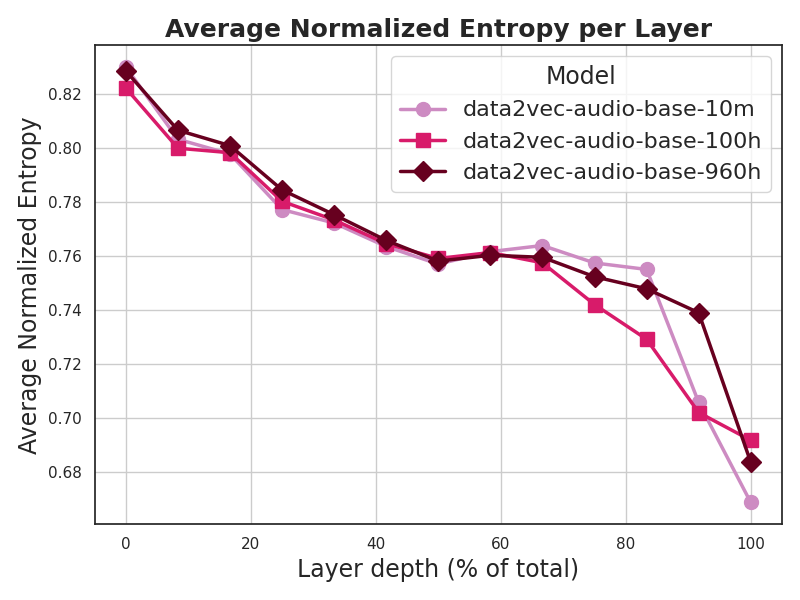}
        \caption{Compression (entropy)}
        \label{fig:finetune-a}
    \end{subfigure}
    \hfill
    \begin{subfigure}[b]{0.32\textwidth}
        \centering
        \includegraphics[width=\textwidth]{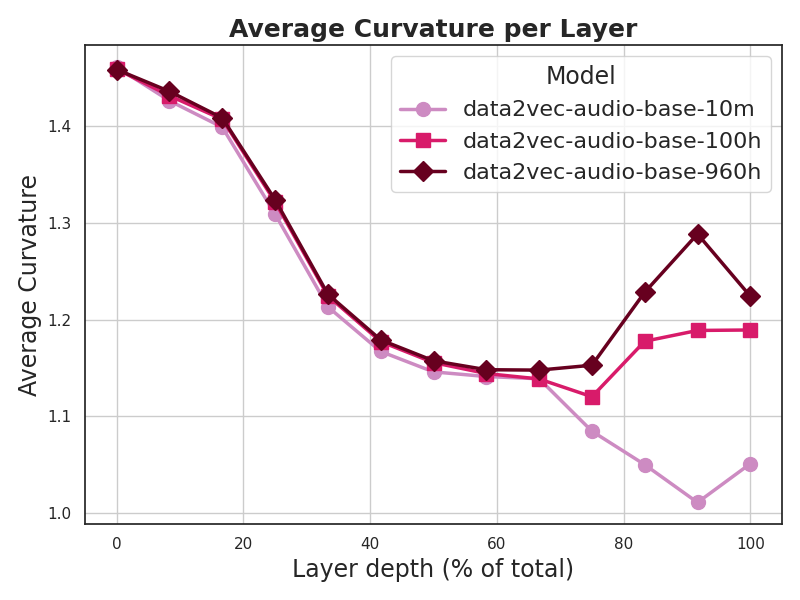}
        \caption{Geometry (curvature)}
        \label{fig:finetune-b}
    \end{subfigure}
    \hfill
    \begin{subfigure}[b]{0.32\textwidth}
        \centering
        \includegraphics[width=\textwidth]{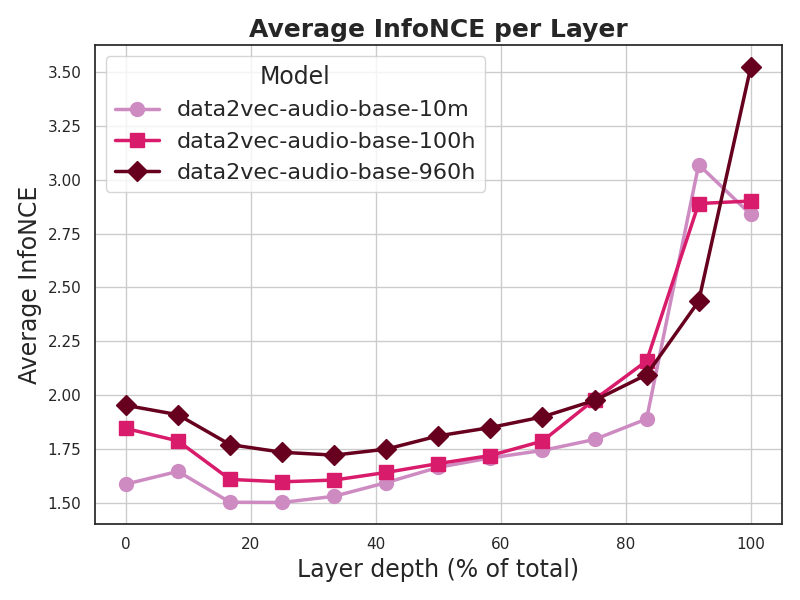}
        \caption{Robustness (InfoNCE)}
        \label{fig:finetune-c}
    \end{subfigure}
    
    \caption{Effect of ASR fine-tuning scale (10\,min, 100\,h, 960\,h) on \ssl{data2vec} representations, illustrating how layer-wise properties (compression, geometry, and robustness) evolve after task adaptation.}
    \label{fig:finetune}
\end{figure*}

\subsection{Per-Layer Analysis: Results \& Discussion}
\label{sec:analysis}

Figure~\ref{fig:analysis-3-perspectives} presents a systematic evaluation of the internal dynamics of different representative SSL models. The top row (Figs.~\ref{fig:fig1a}--\ref{fig:fig1c}) tracks the layer-wise evolution of the three core metrics—entropy, curvature, and invariance. The \ssl{MelSpec} baseline is included as a reference for raw acoustic features. 
Complementing these trajectories, the bottom row (Figs.~\ref{fig:fig1d}--\ref{fig:fig1f}) displays inter-model correlation matrices. These heatmaps quantify the pairwise similarity of layer-wise behaviors between architectures using \textit{Pearson} correlation.\\

\noindent\textbf{Information-theoretic perspective.}
Figure~\ref{fig:fig1a} illustrates the layer-wise evolution of average normalized entropy across audio SSL models.
The \ssl{MelSpec} (dashed line) baseline remains distinctively low due to inherent signal redundancy, which concentrates variance into a few dominant axes. This results in a "spiky" eigenvalue distribution that naturally minimizes entropy.
In contrast, most architectures (e.g., \ssl{WavLM}, \ssl{HuBERT}, \ssl{Data2vec}) sustain high entropy levels throughout the network, initiating at an average of $\approx 0.82$ and exhibiting only a modest decline to $\approx 0.75$ in the final layers. This steady trajectory indicates that informational density is consistently preserved rather than aggressively compressed. These high values imply that the representations maintain a rich utilization of the latent space (high effective rank), avoiding collapse onto a low-dimensional manifold.
Crucially, \ssl{Wav2Vec2-base} (blue curve) diverges from this pattern: while it aligns with other models in early layers, it undergoes an \emph{entropy collapse} towards the end of the Transformer layers (at layer~11). In these final layers, entropy drops to near-baseline levels. This suggests a compression of representational diversity just before the output. While intuitively this might be linked to Wav2Vec2's discrete quantization module used during pretraining, other factors such as specific optimization dynamics, architectural nuances, or the influence of projection heads cannot be ruled out.
The inter-model entropy correlation matrix (Figure~\ref{fig:fig1d}) confirms the consistency of these dynamics across the other architectures. Specifically, we observe high entropy correlations (exhibiting an average correlation of $0.86$) among \ssl{HuBERT}, \ssl{WavLM}, and \ssl{UniSpeech}, indicating shared trajectories in information compression.\\

\noindent\textbf{Geometric perspective.}
Figure~\ref{fig:fig1b} reports the average curvature of the representation manifold across layers, serving as a proxy for the linearity of the data space \cite{hosseini2023large}. 
Most models (e.g., \ssl{WavLM}, \ssl{HuBERT}) exhibit two main regimes: curvature remains initially high ($\approx 1.4$) in the early layers, reflecting feature complexification, and progressively decreases to stabilize around $1.2$ in the deeper layers.
This trajectory suggests that the network “unfolds” the representation manifold to facilitate linear separability.
Relative to this dynamic, the \ssl{MelSpec} baseline serves as a reference for minimal curvature, indicating that raw input features occupy a flatter geometric space than the learned representations.
Distinct behaviors are observed in other architectures: \ssl{data2vec} (red curve) starts with the highest initial curvature and reduces it almost linearly to below-baseline levels, while \ssl{Wav2Vec2} displays a late-stage fluctuation around the 80\% depth mark, consistent with the entropy collapse observed in Figure~\ref{fig:fig1a}.
Similarly, the inter-model curvature correlation matrix (Figure~\ref{fig:fig1e}) demonstrates an even stronger consistency in geometric evolution. Specifically, we observe high curvature correlations (exhibiting a correlation superior to 0.96) among \ssl{HuBERT}, \ssl{WavLM}, and \ssl{UniSpeech}, indicating that these models share a nearly identical strategy for manifold unfolding.\\

\noindent\textbf{Robustness perspective.}
Figure~\ref{fig:fig1c} tracks the average \textit{InfoNCE} loss, serving as a proxy for assessing the robustness/invariance of SSL models to various perturbations (see Section~\ref{subsec:setup}).
Most architectures (e.g., \ssl{HuBERT}, \ssl{WavLM}) rapidly minimize this loss within the first $20$\% of layers, maintaining a stable low plateau that signifies sustained discriminative robustness.
In sharp contrast, \ssl{Wav2Vec2-base} (blue curve) and \ssl{data2vec} (red curve) diverge in the deep layers: after an initial minimization, both models exhibit a sudden spike in InfoNCE near $90$\% depth.
For \ssl{Wav2Vec2}, this degradation aligns with the entropy collapse observed in Figure~\ref{fig:fig1a}. This suggests a regime shift where discriminative performance drops near the output, a phenomenon that may be tied to vector quantization, projection heads, or specific optimization dynamics.
Finally, the inter-model invariance correlation matrix (Figure~\ref{fig:fig1f}) validates this behavioral separation. The analysis identifies two clear clusters: a stable group comprising \ssl{HuBERT}, \ssl{UniSpeech}, and \ssl{WavLM}, which exhibit strong mutual correlations consistent with their sustained invariance; and a second group formed by \ssl{Wav2Vec2} and \ssl{data2vec}. The high correlation between these latter two confirms that their late-stage deviation is intrinsic to their shared optimization regime, distinguishing them from the HuBERT-style masked prediction models.

\begin{tcolorbox}[
    enhanced,
    frame hidden,          
    colback=gray!5,        
    borderline west={2pt}{0pt}{gray!50}, 
    sharp corners,         
    boxsep=0pt,
    left=10pt, right=5pt, top=5pt, bottom=5pt 
]
    \small
    \textbf{Take-away: Complexity $\to$ Abstraction.} \\
    SSL audio models first increase feature complexity (curvature), then unfold their embedding manifold, and finally stabilize in deeper layers.
\end{tcolorbox}

For a comprehensive and interactive exploration of our layer-wise results using the model-centric perspective across these and other models, we encourage readers to visit the \href{https://samsad35.github.io/audio-ssl-dynamics-site/}{\texttt{project website}}.

\subsection{Impact of Scale, Data, Training and Robustness}
\noindent\textbf{Impact of scale and data.} Figure~\ref{fig:size} examines how model scale and training data volume influence internal representations, focusing on different \ssl{WavLM} configurations analyzed in Figure 3. The results show that scaling model size has a stronger structural effect than simply increasing the amount of training data: the trajectories of \ssl{wavlm-base} and \ssl{wavlm-base-plus} are closely aligned, whereas \ssl{wavlm-large} exhibits a markedly different behavior.
In terms of entropy, \ssl{wavlm-large} exhibits a distinct trajectory: starting at a value of $0.70$ in layer $1$, it rapidly climbs to $0.80$ by the 30\% depth mark. This initial low entropy reflects a strong compression of the representation into a specialized subspace. From a geometric perspective, the model resists the late-stage linearization observed in smaller models, preserving richer manifold structures even in the deepest layers. Finally, the InfoNCE analysis highlights divergent optimization strategies: base models quickly achieve and maintain high discriminative performance (low loss) by mid-network, while the large model begins with much higher loss and only reaches optimal discrimination in the final layers.

\begin{figure}[t!]
    \centering
    \begin{subfigure}[b]{0.42\textwidth}
        \centering
        \includegraphics[width=\textwidth]{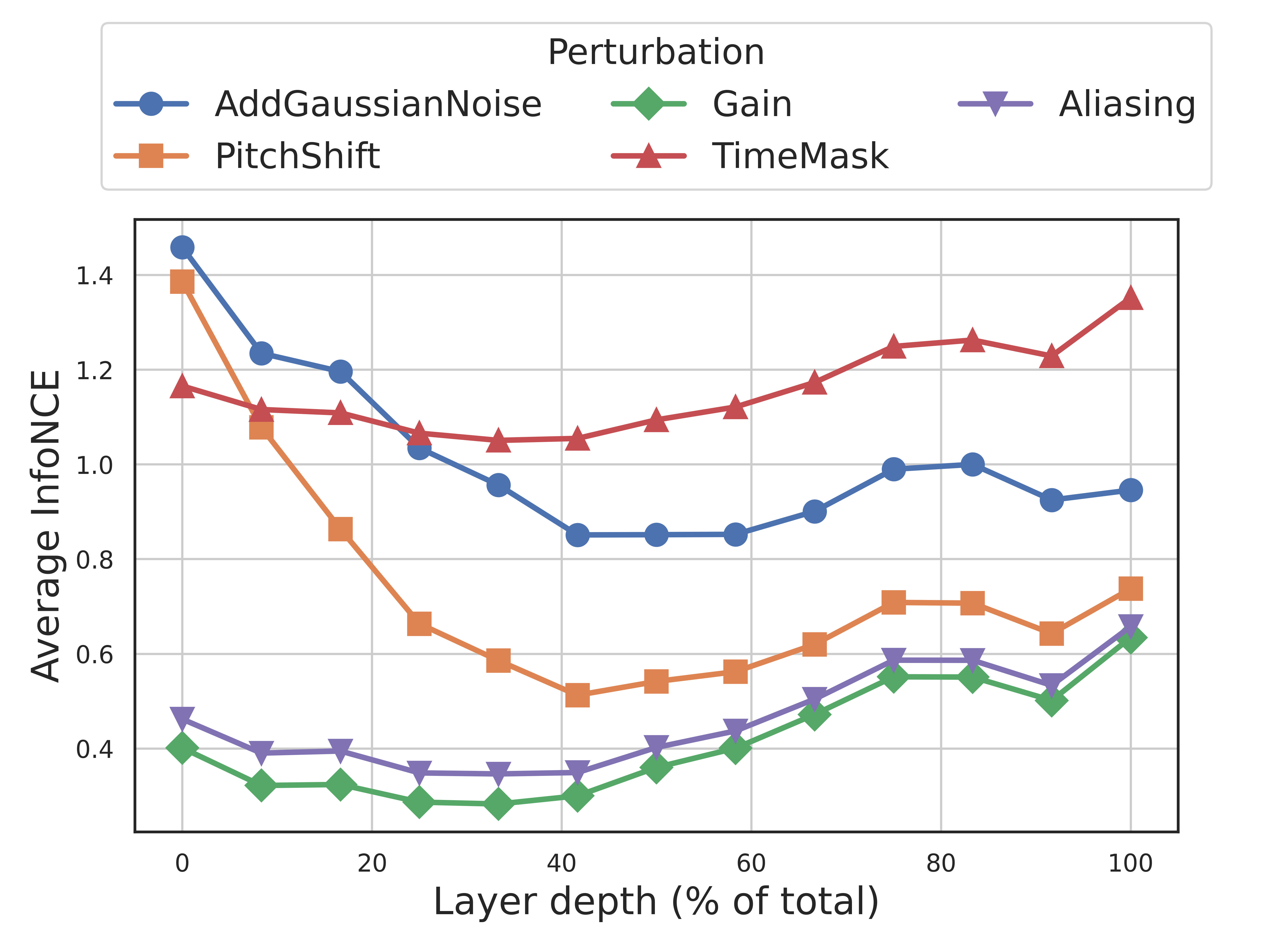}
        \caption{\ssl{WavLM}}
        \label{fig:robustness-wavlm}
    \end{subfigure}
    \hfill
    \begin{subfigure}[b]{0.42\textwidth}
        \centering
        \includegraphics[width=\textwidth]{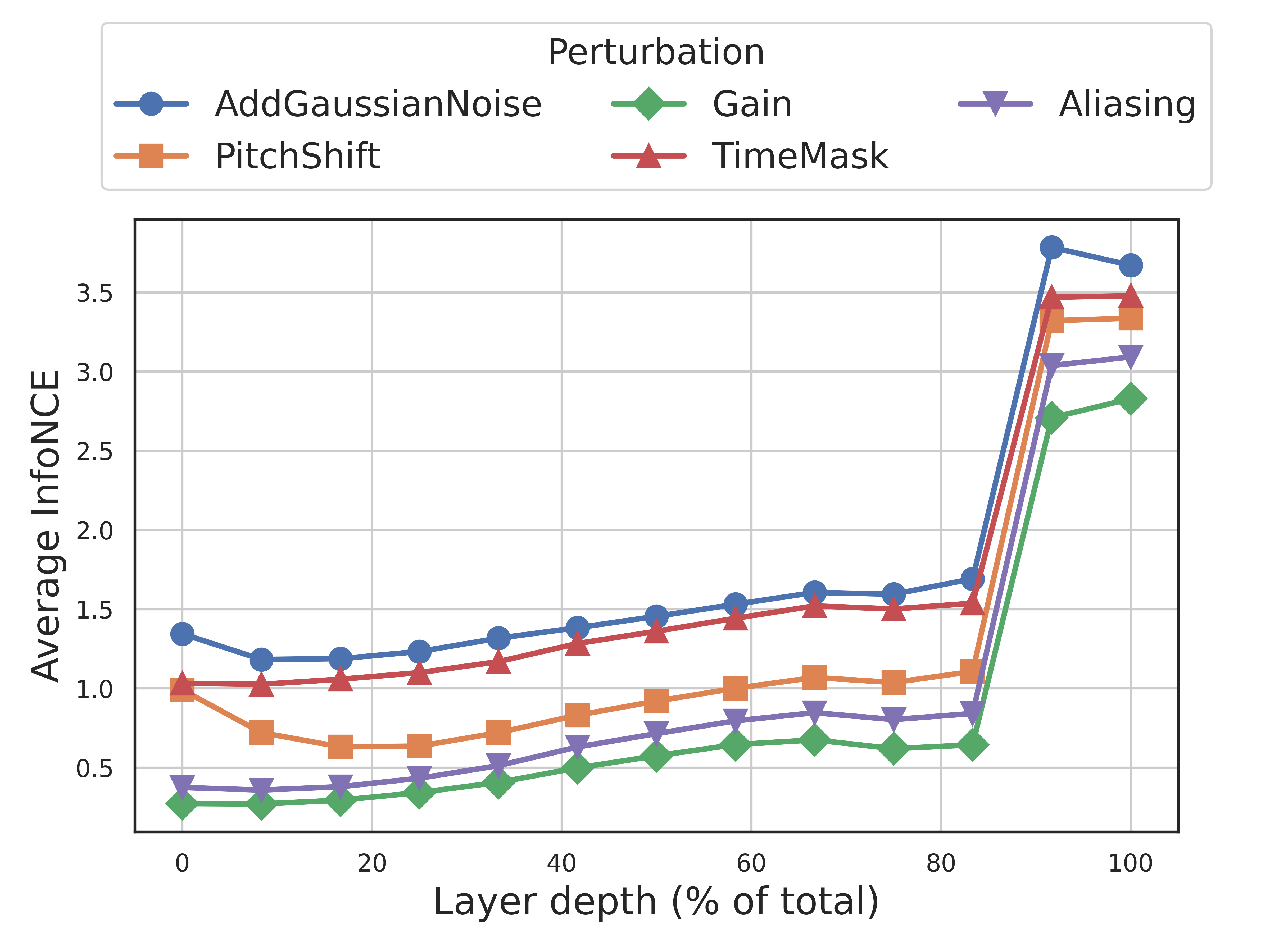}
        \caption{\ssl{Wav2Vec2}}
        \label{fig:robustness-wav2vec}
    \end{subfigure}
    \caption{Robustness analysis--Layer-wise impact of diverse perturbations on representation invariance. We report Average InfoNCE for \ssl{WavLM} (a) and \ssl{Wav2Vec2} (b) across layers.}
    \label{fig:robustness}
\end{figure}


\noindent\textbf{Impact of fine-tuning data scale.~} Figure~\ref{fig:finetune} investigates how the volume of labeled data during ASR fine-tuning ($10$\,min vs. $100$\,h vs. $960$\,h) reshapes the internal representations of \ssl{data2vec}.
Remarkably, entropy (Figure~\ref{fig:finetune-a}) exhibits negligible variation across data regimes. This stability implies that the informational density and effective dimensionality of the representations are determined primarily by pre-training, remaining invariant to the scale of downstream supervision.
In contrast, curvature reveals a two-phase evolution. From layers $0$ to $8$, the trajectories remain tightly aligned across all regimes, indicating that early-stage feature extraction is invariant to fine-tuning data volume. However, starting at layer $9$, a clear distinction emerges. As the volume of fine-tuning data increases (from $10$m to $960$h), the curvature progressively rises, thereby deviating further from the low-curvature regime characteristic of the pre-trained state in deep layers. This suggests that extensive supervision compels the model to construct a more geometrically complex representation manifold.
Furthermore, the InfoNCE loss increases with depth for all models, with the $960$h model exhibiting the highest final loss. This pattern indicates that extensive fine-tuning gradually overrides the original contrastive alignment, shifting the model away from its pre-training objective to specialize more effectively for the downstream task.
    

\noindent\textbf{Robustness analysis.}
Figure~\ref{fig:robustness} illustrates the invariance analysis, evaluating \ssl{WavLM} (\ref{fig:robustness-wavlm}) and \ssl{Wav2Vec2} (\ref{fig:robustness-wav2vec}) representations under specific, individual signal perturbations rather than a global aggregation as in Figure~\ref{fig:fig1d}.
Globally, \ssl{WavLM} consistently outperforms \ssl{Wav2Vec2}, maintaining lower InfoNCE loss across most perturbation categories. This superior invariance aligns with the \ssl{WavLM} training objective, which integrates a denoising masked prediction task, effectively regularizing the model against acoustic distortions.
Consequently, \ssl{WavLM} demonstrates a progressive emergence of invariance, particularly regarding additive and spectral distortions like \textit{AddGaussianNoise} (blue) and \textit{PitchShift} (orange). In these cases, the initially high sensitivity significantly decreases in the deeper layers, indicating that the model constructs representations that are increasingly robust to these artifacts.
By contrast, \textit{TimeMask} (red) poses a persistent challenge for both architectures: its loss remains high and flat, suggesting that models struggle to recover information explicitly removed from the temporal domain, as opposed to merely distorted signals.
Finally, a distinct behavior emerges in the deep layers of \ssl{Wav2Vec2}. We consistently observe a sharp rebound in InfoNCE loss across the final two layers for all perturbations, increasing from an average of $1.0$ to $3.0$. This suggests that while middle layers build invariance, the model's output layers reintroduce sensitivity to acoustic variations, likely to satisfy the fine-grained requirements of its quantization objective.

\begin{figure}[t!]
    \centering
    \includegraphics[width=0.80\linewidth]{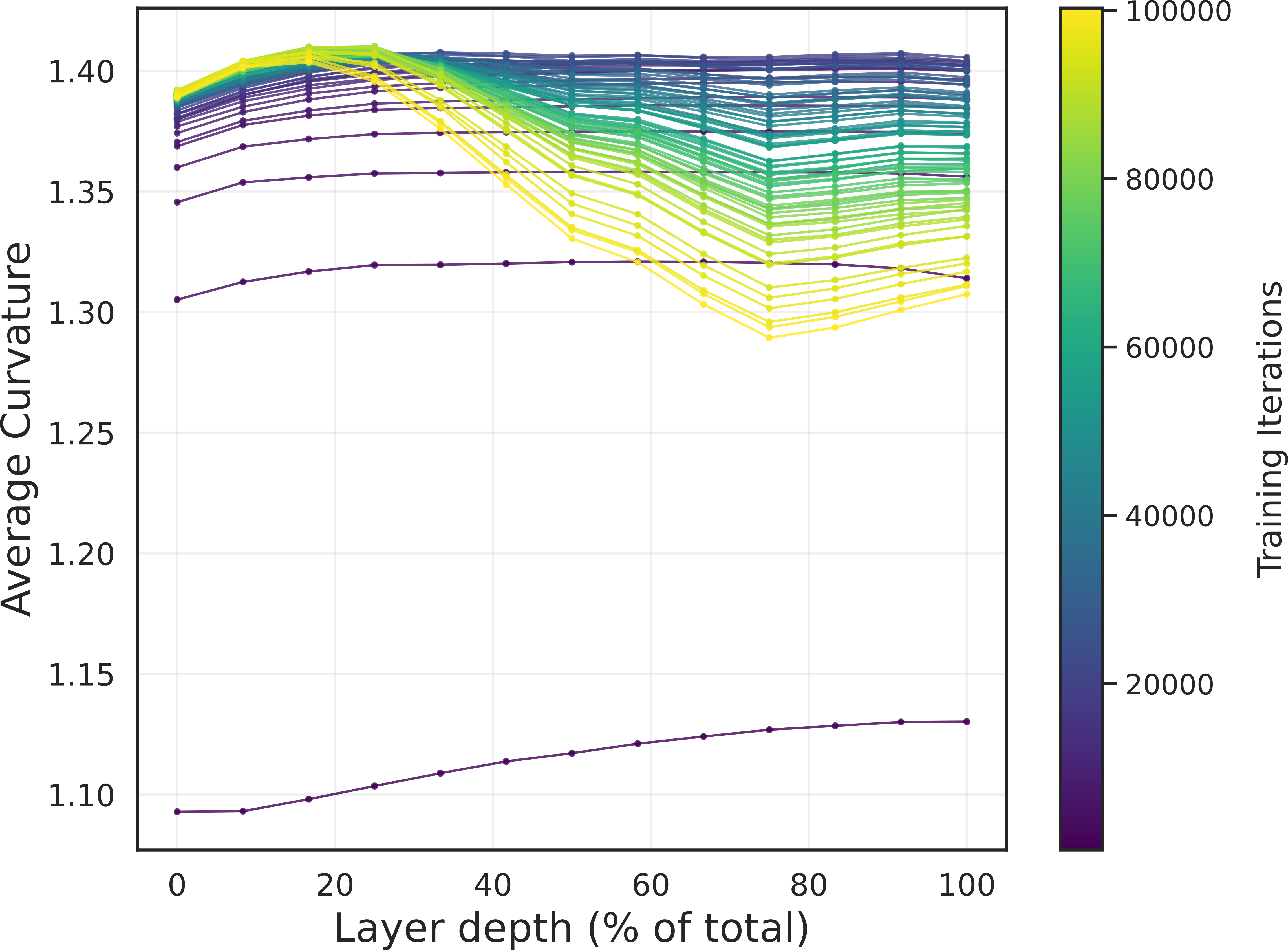}
    \caption{Evolution of layer-wise curvature during training. The curves display the average curvature profile across the network depth (normalized $0$--$100\%$), colored by training iterations (from purple: initialization, to yellow: $100$k iterations).}
    \label{fig:curvature-evolution}
\end{figure}

\noindent\textbf{Training dynamics and geometric relaxation.~} Figure~\ref{fig:curvature-evolution} tracks the evolution of curvature during the first training iteration of a \ssl{HuBERT} model, utilizing labels extracted from MFCCs. The color gradient shifts from initialization (purple) to convergence (yellow). Initially, the model exhibits a low and relatively uniform curvature ($\approx 1.16$) across all layers. Very quickly, as optimization begins (dark purple to teal), we observe a sharp global surge, with curvature values rising to approximately $1.42$, reflecting the rapid encoding of acoustic complexity. As training progresses (green to yellow), a clear structural separation emerges. While the early layers ($0$--$40\%$) maintain high curvature to encode intricate features, the deeper layers ($>40\%$) undergo a progressive relaxation. In this final phase, the curvature systematically drops, signaling a linearization process where the model actively ``flattens'' its deep representations to facilitate linear separability.

\begin{figure*}[t!]
    \centering
    \begin{subfigure}[b]{0.24\textwidth}
        \centering
        \includegraphics[width=\textwidth]{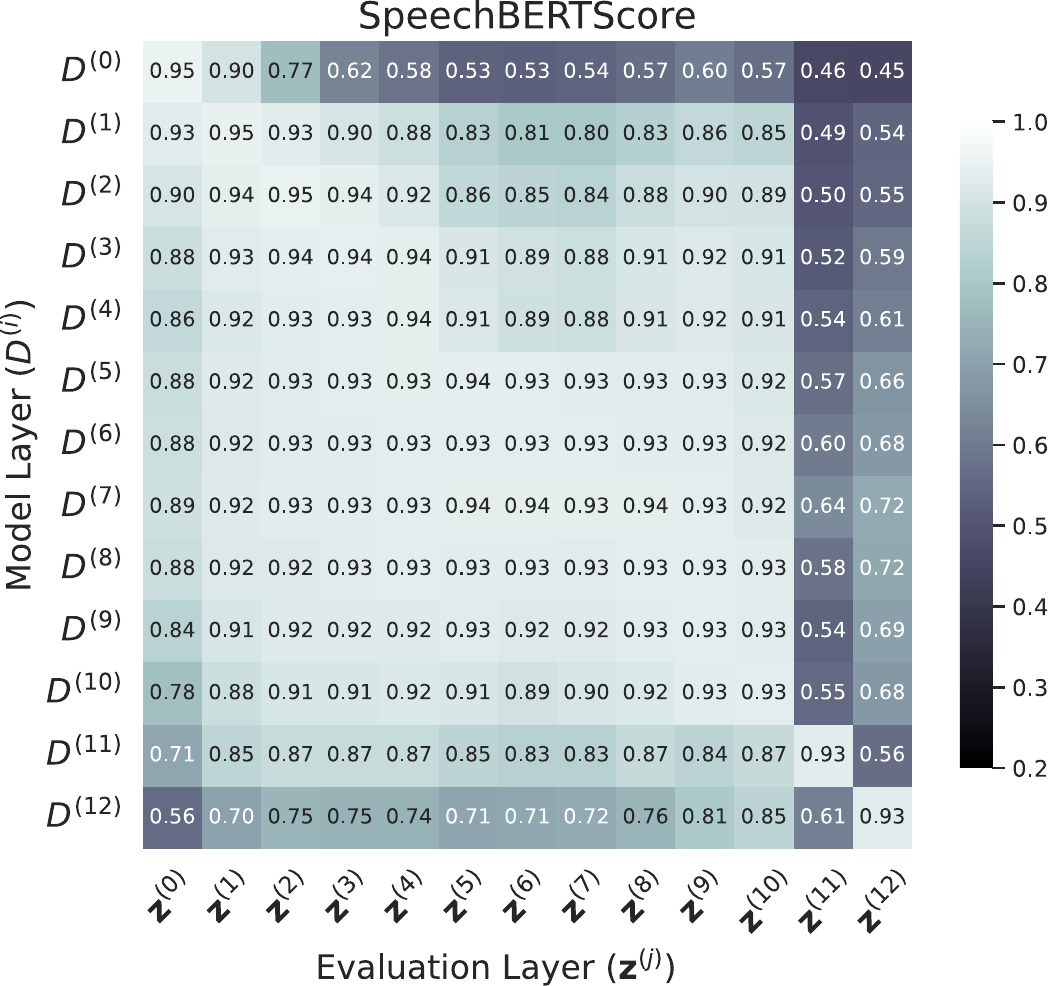}
        \caption{\ssl{Wav2Vec2}: content-GCM}
        \label{fig:clpm-content-w2v}
    \end{subfigure}
    \hfill
    \begin{subfigure}[b]{0.24\textwidth}
        \centering
        \includegraphics[width=\textwidth]{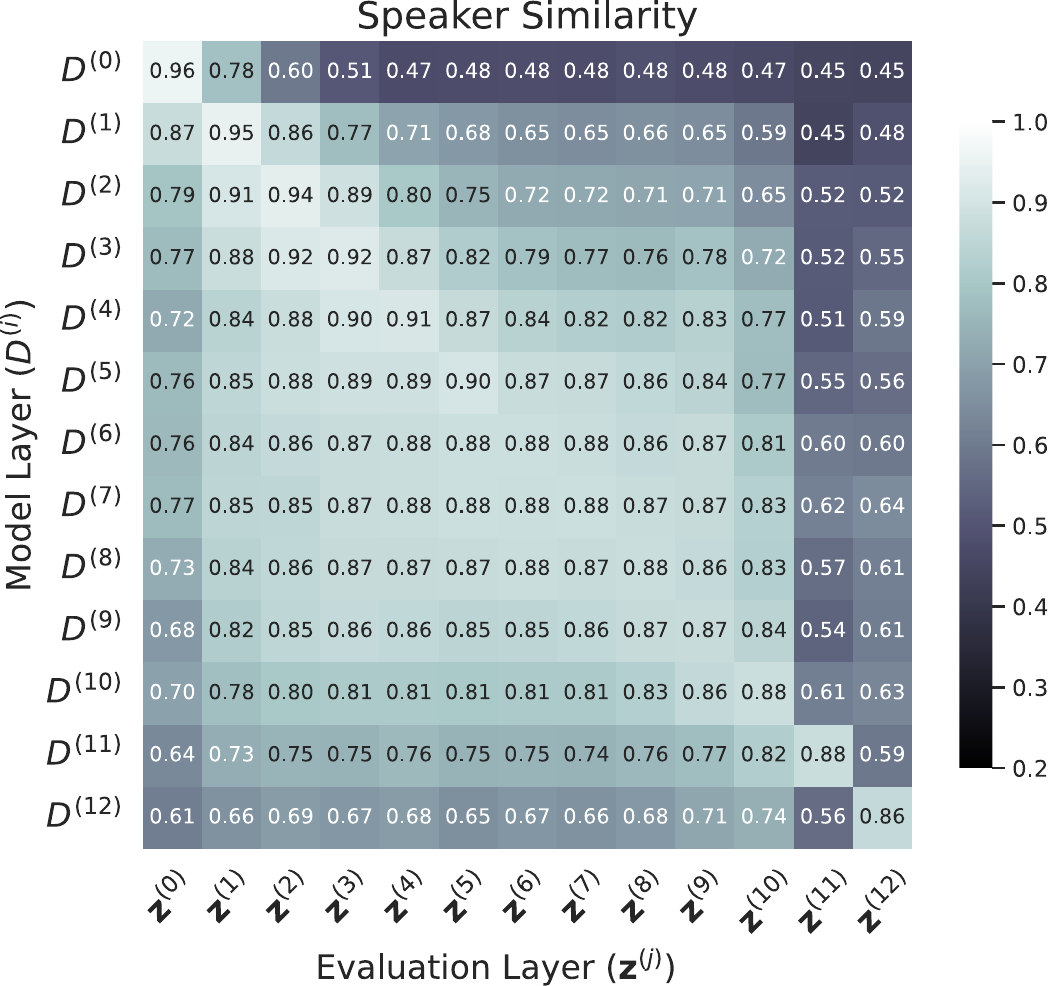}
        \caption{\ssl{Wav2Vec2}: speaker-GCM}
        \label{fig:clpm-speaker-w2v}
    \end{subfigure}
    \hfill
    \begin{subfigure}[b]{0.24\textwidth}
        \centering
        \includegraphics[width=\textwidth]{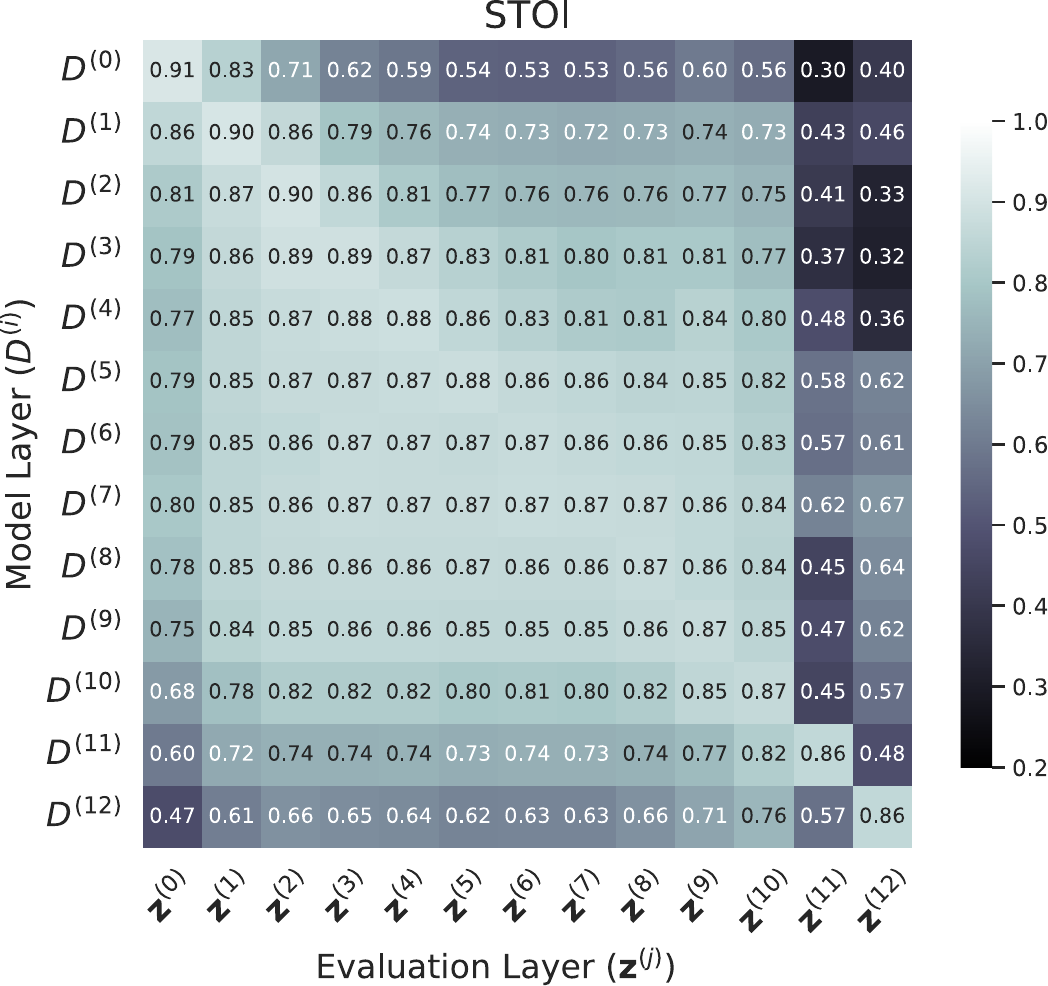}
        \caption{\ssl{Wav2Vec2}: STOI-GCM}
        \label{fig:clpm-stoi-w2v}
    \end{subfigure}
    \hfill
    \begin{subfigure}[b]{0.24\textwidth}
        \centering
        \includegraphics[width=\textwidth]{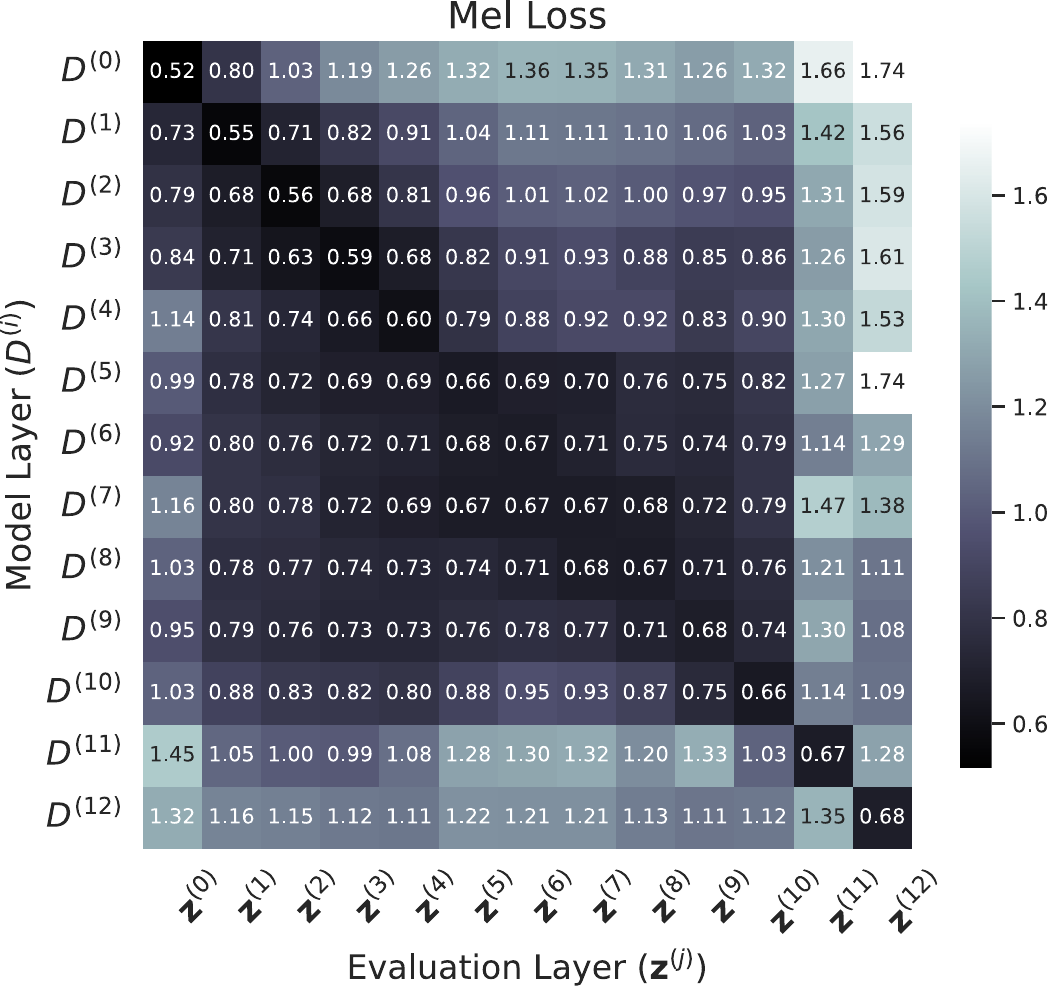}
        \caption{\ssl{Wav2Vec2}: Mel Recons.}
        \label{fig:clpm-mel-wavlm}
    \end{subfigure}

    \vspace{0.3cm} 

    \begin{subfigure}[b]{0.24\textwidth}
        \centering
        \includegraphics[width=\textwidth]{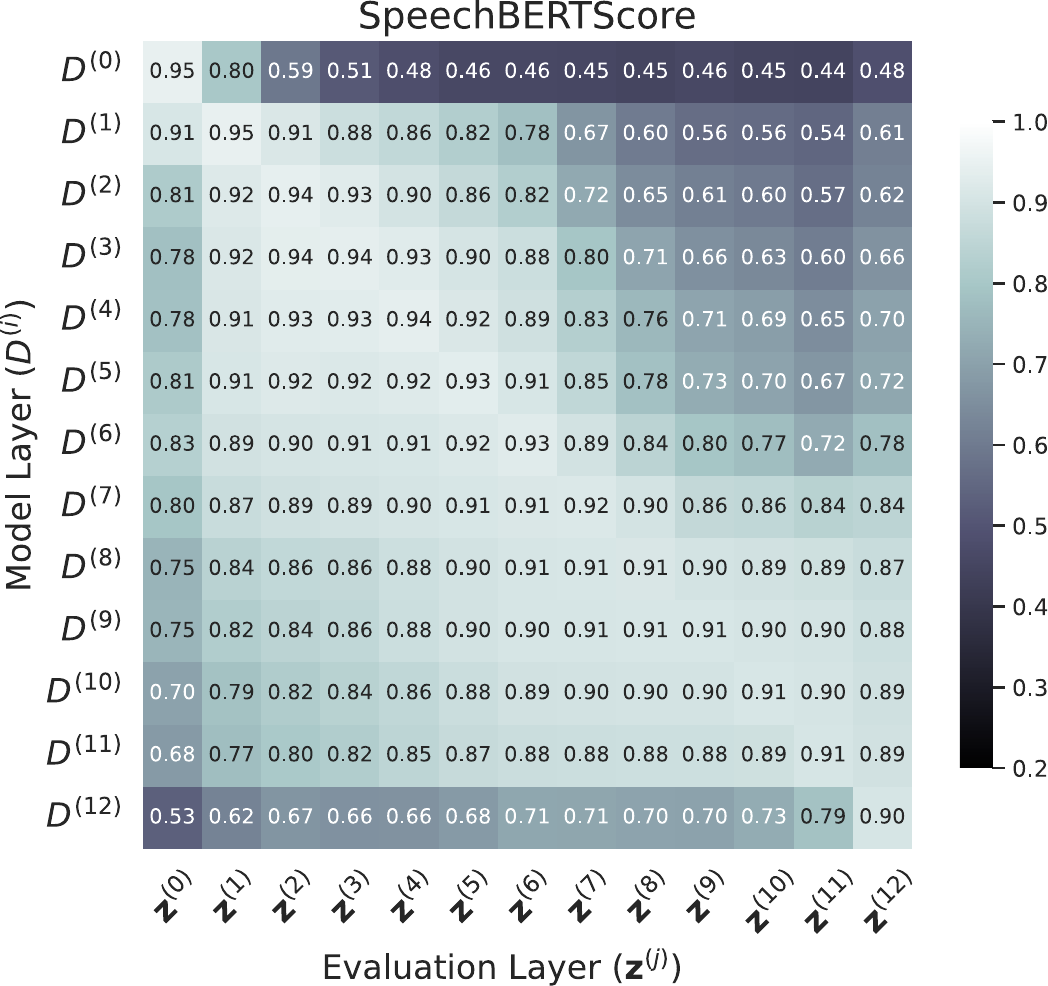}
        \caption{\ssl{WavLM}: content-GCM}
        \label{fig:clpm-content-wavlm}
    \end{subfigure}
    \hfill
    \begin{subfigure}[b]{0.24\textwidth}
        \centering
        \includegraphics[width=\textwidth]{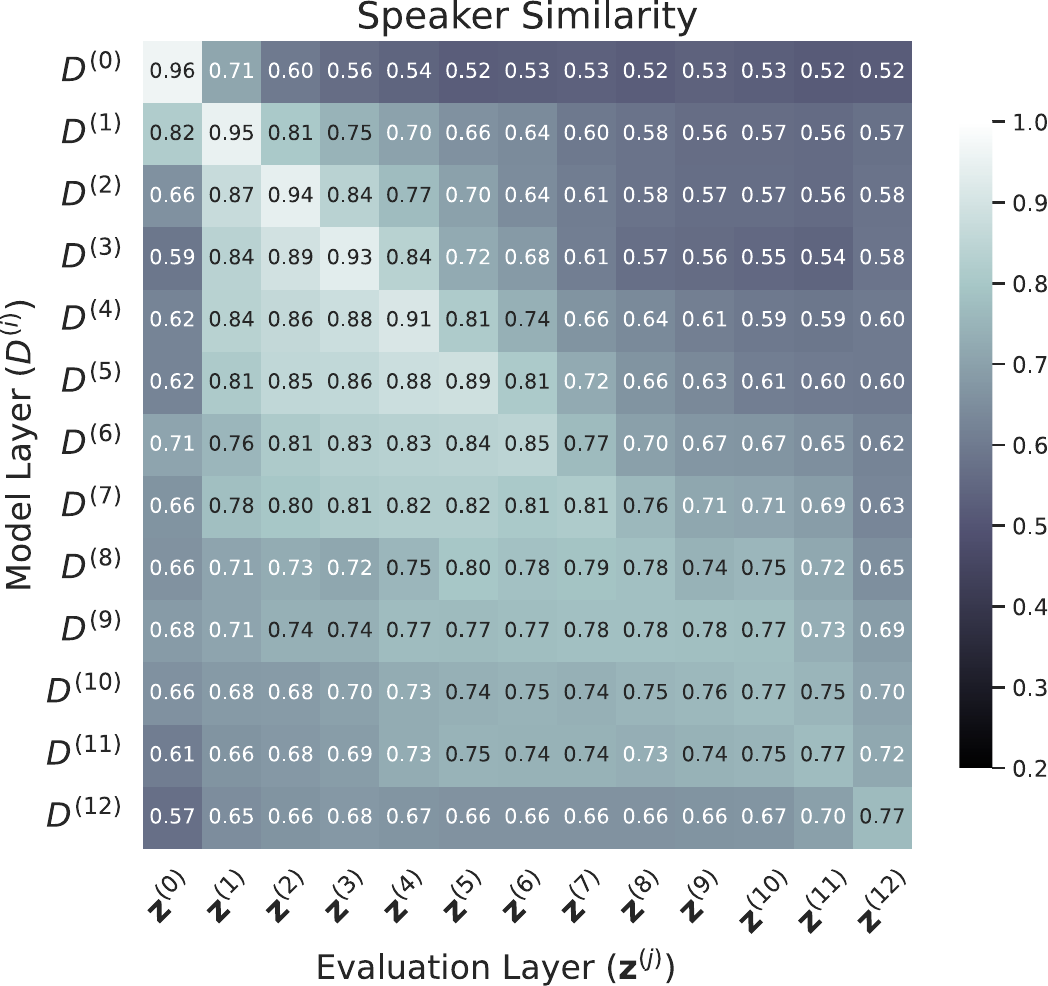}
        \caption{\ssl{WavLM}: speaker-GCM}
        \label{fig:clpm-speaker-wavlm}
    \end{subfigure}
    \hfill
    \begin{subfigure}[b]{0.24\textwidth}
        \centering
        \includegraphics[width=\textwidth]{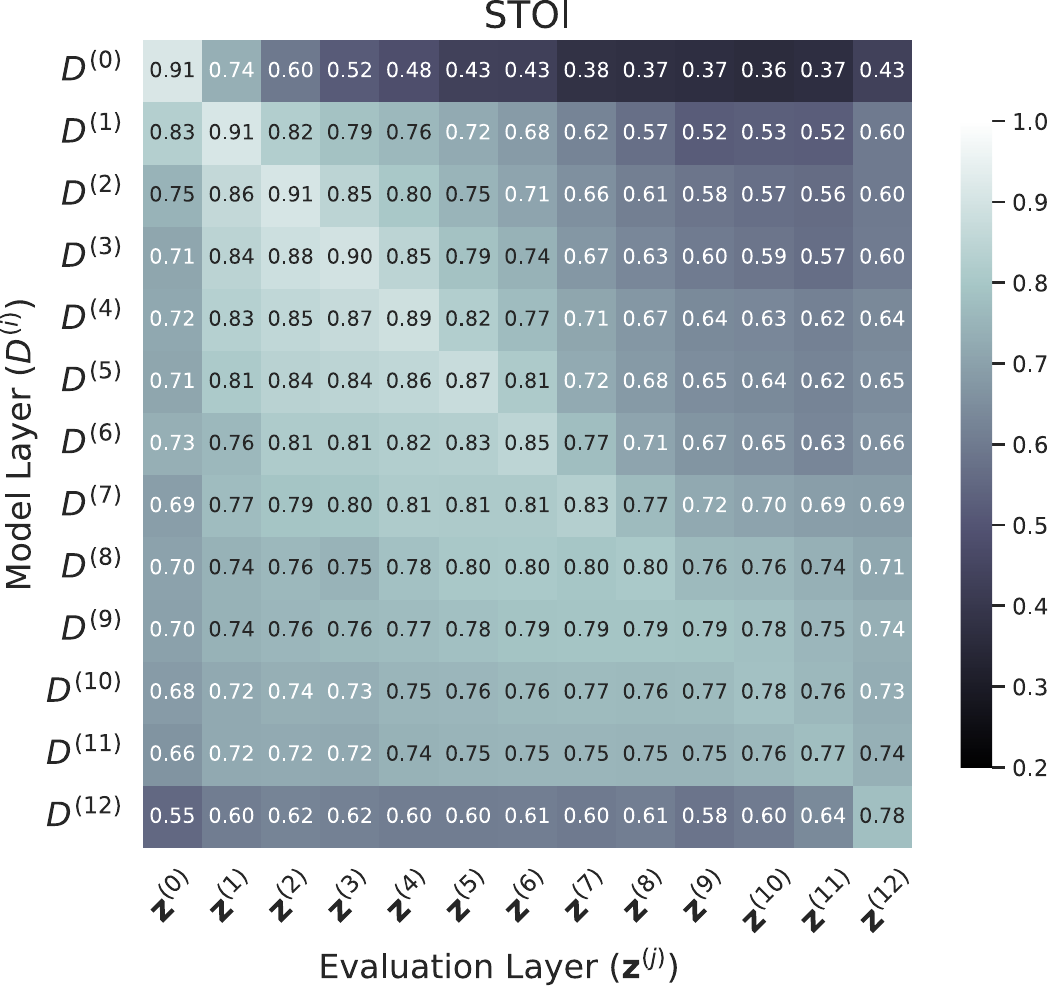}
        \caption{\ssl{WavLM}: STOI-GCM}
        \label{fig:clpm-stoi-wavlm}
    \end{subfigure}
    \hfill
    \begin{subfigure}[b]{0.24\textwidth}
        \centering
        \includegraphics[width=\textwidth]{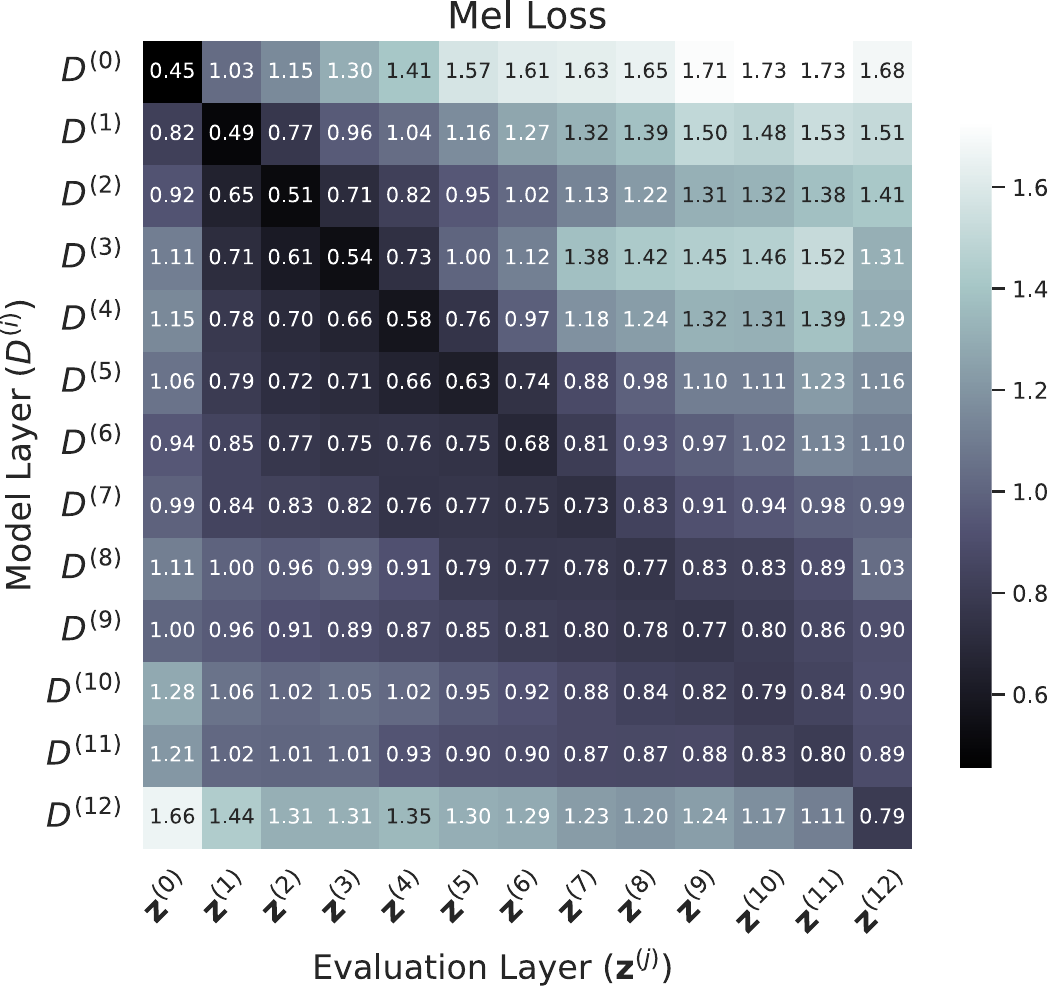}
        \caption{\ssl{WavLM}: Mel Recons.}
        \label{fig:clpm-mel-wavlm}
    \end{subfigure}
    
    \caption{
    The cross-layer Generative Compatibility Matrix (GCM) comparison between Wav2Vec2-base (top) and WavLM-base (bottom). Each cell $(\ell, k)$ represents the performance of a decoder trained on layer $\ell$ and evaluated on layer $k$. 
    }
    \label{fig:GCM-comparison}
\end{figure*}

\subsection{Cross-Layer Analysis, Results \& Discussion}
\noindent\textbf{Experimental setup for the GCM.} We employ a Continuous Flow Matching (CFM) \cite{lipman2022flow} objective to learn the generative mapping from latent SSL embeddings to log Mel-spectrograms. Specifically, the trained decoder $D^{(i)}$ is a 6-layer Diffusion Transformer (DiT) \cite{peebles2023scalable} with a 512 hidden dimension, optimized via CFM. A frozen, pre-trained HiFi-GAN vocoder \cite{kong2020hifi} subsequently reconstructs the final speech waveform without being updated during training. All decoders are trained for 400 epochs on a single GPU using the \emph{train-clean-100} subset, ensuring optimization on high-quality speech. This decoupled configuration (CFM-trained DiT + frozen HiFi-GAN) captures fine-grained acoustics while maintaining sufficient capacity to model the SSL latent space across different depths.

\noindent\textbf{Results and discussion.}
Figure~\ref{fig:GCM-comparison} displays the GCM as heatmaps for \ssl{Wav2Vec2-base} (top row) and \ssl{WavLM-base} (bottom row). We evaluate four metrics $\mathcal{M}$: SpeechBERTScore \cite{saeki2024speechbertscore} for semantic content, Resemblyzer similarity \cite{wan2018generalized} for speaker identity, STOI for audio quality, and $L1$ loss for Mel-spectrogram reconstruction.
In terms of phonetic content (SpeechBERTScore), both SSL models display a broad region of high SpeechBERTScore values, indicating a stable phonetic encoding. For instance, in \ssl{Wav2Vec2}, layers $1$ through $10$ consistently exceed $0.80$ (Figure~\ref{fig:clpm-content-w2v})—a substantial block of linguistic retention. For \ssl{WavLM}, we observe that this stability region is subdivided into two distinct blocks (spanning layers $1$--$6$ and $6$--$12$). This segmentation directly aligns with the two regimes identified in the curvature analysis, a structural pattern that is also consistent with \ssl{Wav2Vec2}. However, in contrast to \ssl{WavLM}, \ssl{Wav2Vec2} exhibits a sharp semantic rupture at layer $11$. This sudden shift directly aligns with the entropy collapse observed in the previous section. Regarding speaker identity, the similarity matrices (Speaker-GCM; Figures~\ref{fig:clpm-speaker-wavlm},~\ref{fig:clpm-speaker-w2v}) also exhibit blocks of correspondence between layers, though these are less pronounced than for content. This indicates that speaker identity is not preserved across distant layers with the same stability.
Furthermore, all matrices are structurally asymmetric ($\text{GCM}(\ell, k) \neq \text{GCM}(k, \ell)$), strongly favoring the lower triangular region. This highlights a strict hierarchical pruning: decoders trained on early, acoustically-rich layers cannot interpret the abstracted representations of deeper layers, whereas deep-layer decoders generalize effectively to preceding ones. 

An interactive audio animation of the GCM evaluations is available on the \href{https://samsad35.github.io/audio-ssl-dynamics-site/#gcm-audio}{\texttt{project website}}.

\begin{figure*}[t!]
    \centering
    \begin{subfigure}[b]{0.32\textwidth}
        \centering
        \includegraphics[width=\textwidth]{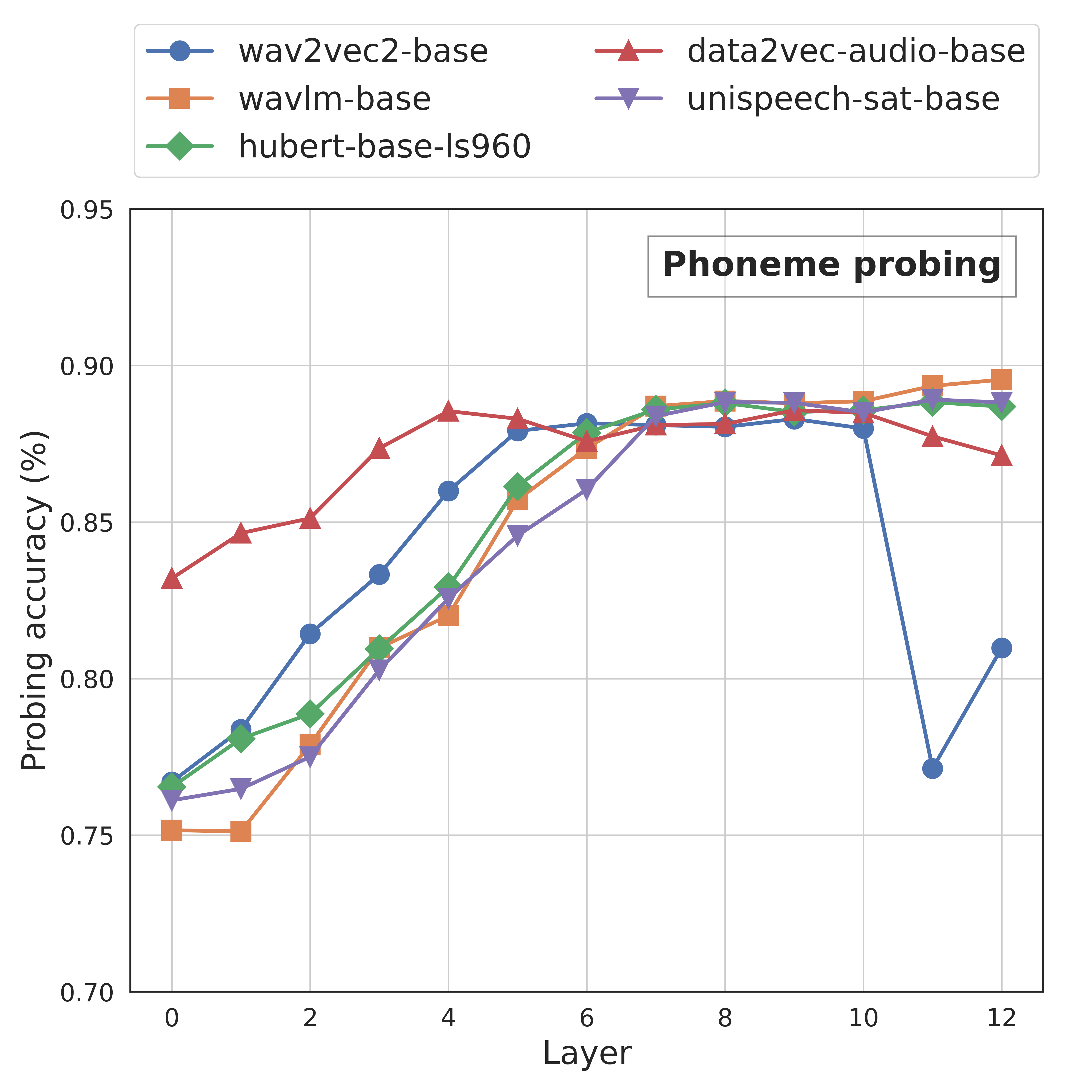}
        \caption{Phoneme classification ($\uparrow$)}
        \label{fig:probe-phoneme}
    \end{subfigure}
    \hfill
    \begin{subfigure}[b]{0.32\textwidth}
        \centering
        \includegraphics[width=\textwidth]{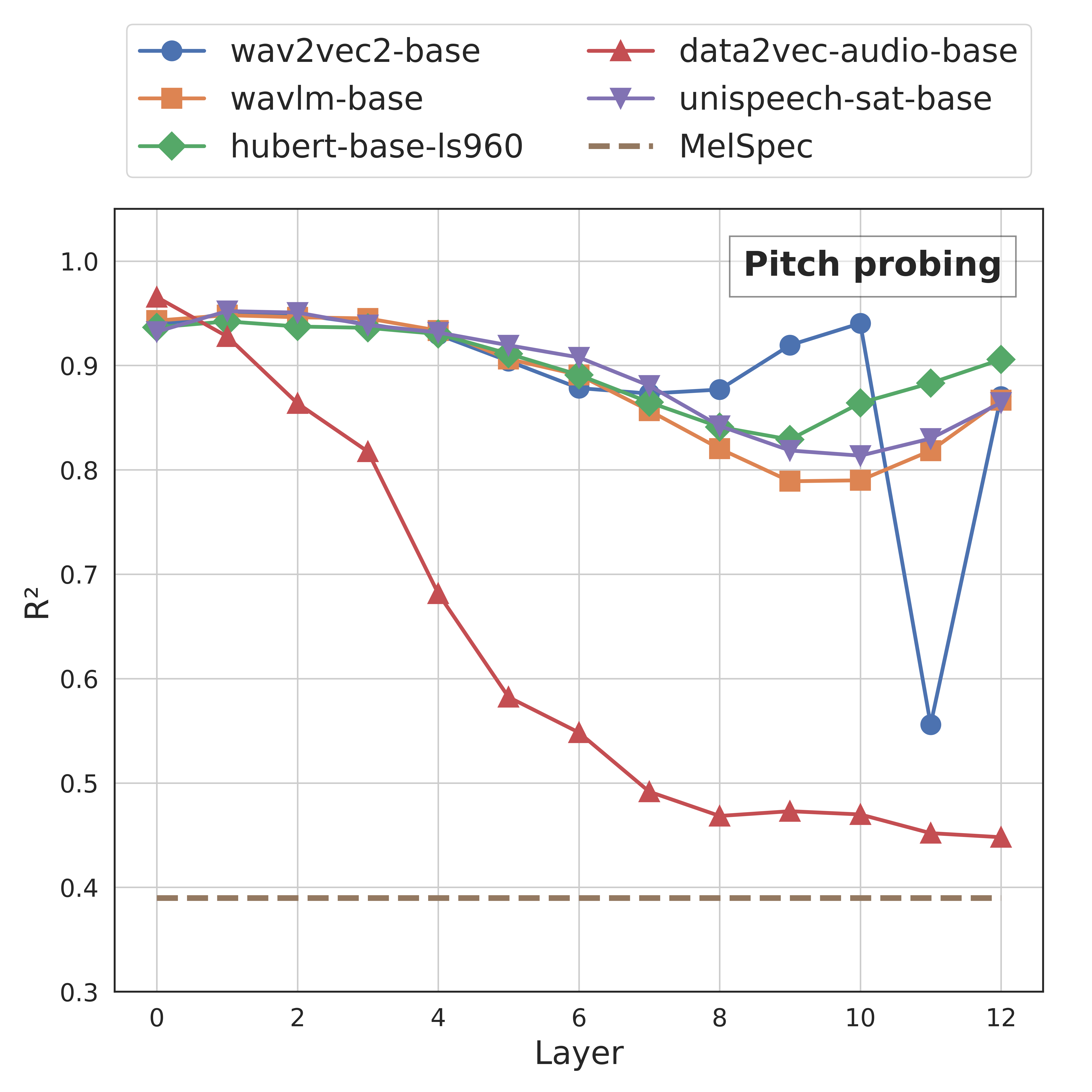}
        \caption{Pitch regression (F0) ($\uparrow$)}
        \label{fig:probe-pitch}
    \end{subfigure}
    \hfill
    \begin{subfigure}[b]{0.32\textwidth}
        \centering
        \includegraphics[width=\textwidth]{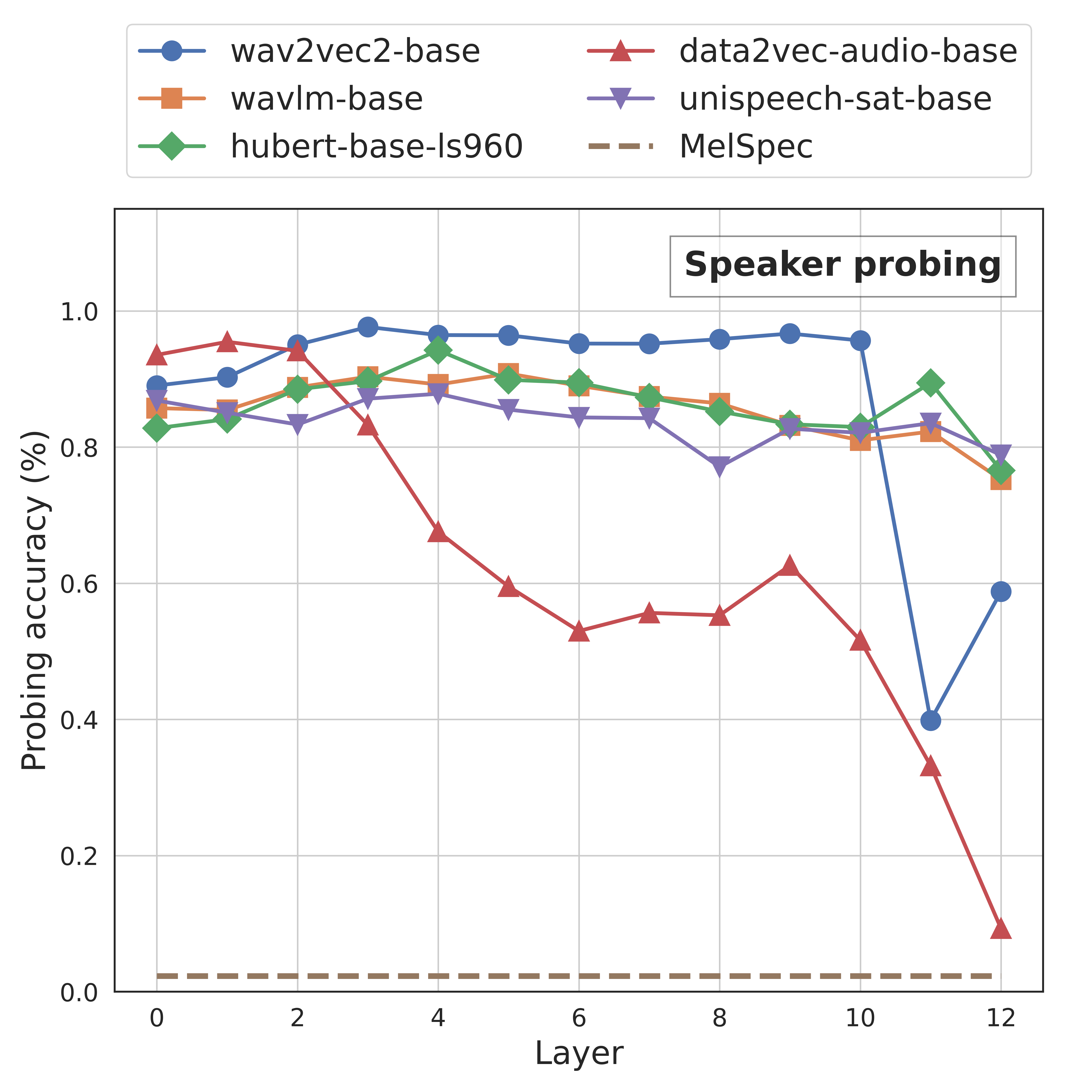}
        \caption{Speaker classification ($\uparrow$)}
        \label{fig:probe-speaker}
    \end{subfigure}
    \caption{Task probing across layers for phoneme classification, pitch regression, and speaker classification. Results show how task-relevant information is distributed across the network hierarchy.}
    \label{fig:task-probing}
\end{figure*}

\begin{figure*}[t!]
    \centering
    \begin{subfigure}[b]{0.32\textwidth}
        \centering
        \includegraphics[width=\textwidth]{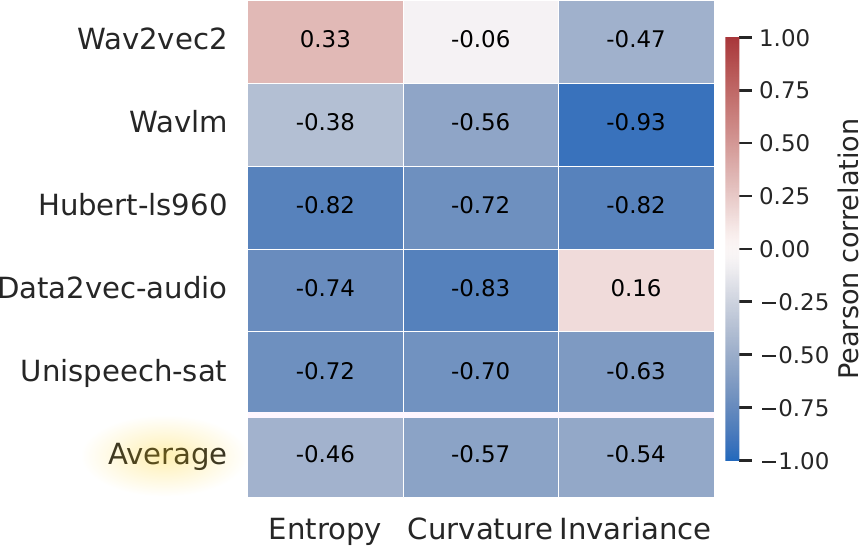}
        \caption{Phoneme classification task}
        \label{fig:corr-phoneme}
    \end{subfigure}
    \hfill
    \begin{subfigure}[b]{0.32\textwidth}
        \centering
        \includegraphics[width=\textwidth]{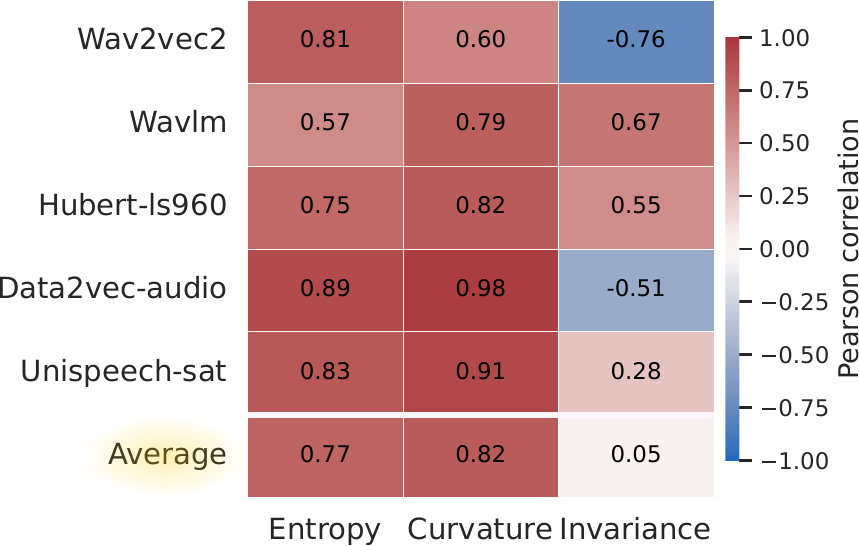}
        \caption{Pitch regression ($F_0$) task}
        \label{fig:corr-pitch}
    \end{subfigure}
    \hfill
    \begin{subfigure}[b]{0.32\textwidth}
        \centering
        \includegraphics[width=\textwidth]{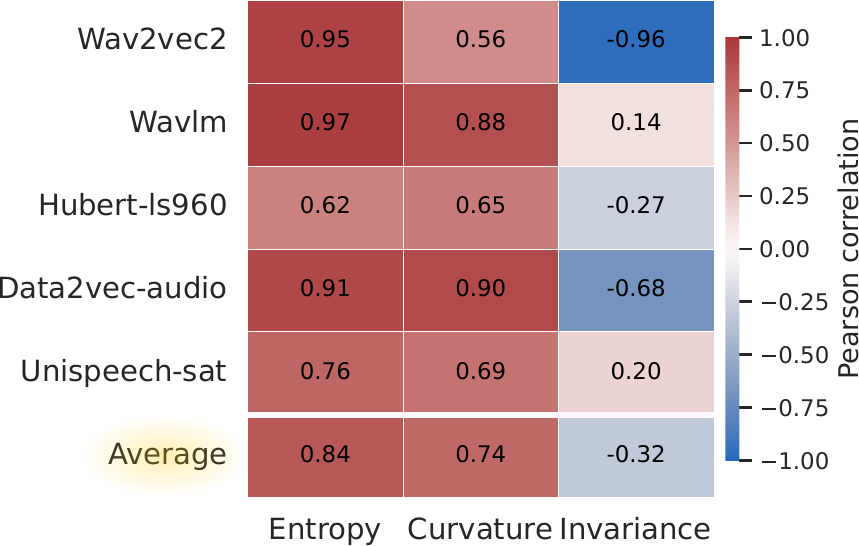}
        \caption{Speaker classification task}
        \label{fig:corr-speaker}
    \end{subfigure}
    \caption{Pearson correlation between layer-wise probing accuracy (PPGs, pitch, speaker identity) and representation properties (entropy, curvature, invariance) across self-supervised speech models. Higher absolute values indicate stronger relationships.}
    \label{fig:task-corr}
\end{figure*}


\subsection{Linking Model-Centric Metrics to Downstream Tasks}
\label{subsec:task-probing}

To connect our intrinsic analyses (\issl) with the downstream task, we conduct linear probing experiments. This involves training linear classifiers or regressors on \emph{frozen SSL} representations to predict task-relevant targets: phoneme identity, speaker identity, or pitch. By restricting the probe to a linear mapping, we quantify the immediate accessibility of information within each layer for downstream tasks.

\noindent\textbf{Probing Setup.} Using the LibriSpeech \textit{train-clean-100} subset, we train linear probes on extracted embeddings $\mathbf{Z}^{(l)}$ to evaluate three distinct tasks: \emph{phoneme classification} to map embeddings to phonetic labels using phonetic posteriorgrams (PPGs \cite{churchwell2024high}), \emph{pitch regression} to predict the fundamental frequency ($F_0$) extracted via CREPE~\cite{kim2018crepe}, and \emph{speaker identification} to classify speaker identities. Probes are trained with early stopping, yielding layer-wise curves that track the evolution of task-specific information across the network hierarchy.

\noindent\textbf{Results and Discussion.} 
Figure~\ref{fig:task-probing} reports linear probing results for {phoneme classification}, {pitch regression}, and {speaker identification}. The curves reveal a clear hierarchy: task-relevant information is not uniformly distributed but concentrated at specific depths, indicating layer-wise specialization for distinct speech attributes.
Phoneme accuracy (Figure~\ref{fig:probe-phoneme}) peaks in mid-layers ($7$--$8$ for \ssl{WavLM}, \ssl{HuBERT}), while \ssl{Data2Vec} peaks earlier at layer $4$. These depths coincide with the curvature transition point (Figure~\ref{fig:fig1b}), marking the shift from a high-curvature encoding regime to a more linearized one. This alignment suggests an optimal trade-off between feature richness and manifold flatness, maximizing phonetic separability before deeper layers over-compress the signal. This layer-wise phonetic specialization is consistent with previous studies \cite{hsu2021HuBERT, pasad2023comparative} reporting that intermediate transformer layers encode the most discriminative phonetic information.
Paralinguistic tasks follow a distinct yet more nuanced pattern. Pitch regression is strongest in the early layers, reflecting the encoding of low-level cues such as fundamental frequency ($F_0$) before abstraction. In models such as \ssl{WavLM} and \ssl{HuBERT}, this information attenuates progressively but remains partially preserved across depth. In contrast, \ssl{Data2Vec} exhibits a sharper decline from around the 3\textsuperscript{rd} layer onward, while \ssl{Wav2Vec2} shows a marked attenuation toward the final layers. Speaker classification closely mirrors these layer-wise patterns, decreasing in deeper layers as terminal representations prune acoustic detail in favor of semantic abstraction. These trends align with prior studies \cite{pasad2023comparative, chiu2025large, chiu2025probing}, which report that paralinguistic information is predominantly encoded in lower layers and progressively diminished as representations become more linguistically specialized.
Pearson correlations in Figure~\ref{fig:task-corr} quantify these trends. Paralinguistic tasks correlate positively with representational complexity: pitch and speaker classification depend strongly on entropy ($0.77, 0.84$) and curvature ($0.82, 0.74$), consistent with a high-dimensional, curved manifold. In contrast, phoneme classification shows negative correlations (Avg: entropy $-0.46$, curvature $-0.57$, invariance $-0.54$), indicating that linguistic discrimination benefits from compression and manifold linearization that suppress raw acoustic variability.

\begin{tcolorbox}[
    enhanced,
    frame hidden,          
    colback=gray!5,        
    borderline west={2pt}{0pt}{gray!50}, 
    sharp corners,         
    boxsep=0pt,
    left=10pt, right=5pt, top=5pt, bottom=5pt 
]
    \small
    \textbf{Take-away: Task Hierarchy.} \\
    Low-level tasks (pitch, speaker) rely on high entropy and curvature, whereas phonemes require deep-layer compression and linearization.
\end{tcolorbox}

\section{Conclusion}

In this work, we introduced the \issl\ framework, a unified, \emph{task-agnostic and model-centric} approach to analyze SSL speech representations first through three per-layer lenses: compression, geometry, and robustness. Our analysis revealed distinct optimization regimes, notably the late-stage \textit{entropy collapse} in Wav2Vec2, contrasting with the geometric stability of WavLM. Second, we presented a novel method to estimate functional compatibility across the network hierarchy, the cross-layer Generative Compatibility Matrix (GCM), which uncovered a stable phonetic core in mid-layers. In addition, we explicitly linked these intrinsic properties to downstream performance via task probing. This confirmed that phoneme recognition benefits from deep-layer compression, whereas pitch and speaker tasks rely on early high-entropy states, offering a roadmap for designing efficient, task-aligned architectures. While our framework provides a robust empirical foundation, future work must establish formal causal links. Isolating the precise mechanisms behind phenomena like Wav2Vec2's extreme deep-layer compression will be crucial for designing next-generation models.

\section{Acknowledgments}
This work was granted access to the HPC resources of IDRIS under the allocation Grant 2025-A0181016041 made by GENCI.

\section{Generative AI Use Disclosure}
Generative AI tools were used exclusively for language editing and stylistic improvements. They did not contribute to the scientific content, analyses, or conclusions of this work. All authors take full responsibility for the manuscript, have approved its submission, and confirm that no generative AI system is listed as a co-author, in accordance with ISCA policy.

\bibliographystyle{IEEEtran}
\bibliography{mybib}

\end{document}